\documentclass[10pt,journal,compsoc]{IEEEtran}

\usepackage{ctable}
\usepackage{caption}

\ifCLASSOPTIONcompsoc
  \usepackage[nocompress]{cite}
\else
  \usepackage{cite}
\fi

\usepackage{graphicx}
\usepackage{float}
\usepackage{amssymb,amsfonts,amsmath,amsthm}
\DeclareMathOperator{\sign}{sign}

\usepackage{breqn}
\usepackage{algorithm,algpseudocode}
\usepackage{enumitem}
\usepackage[hidelinks]{hyperref}
\usepackage{multirow}
\begin{document}

\title{A Recommendation Approach based on Similarity-Popularity Models of Complex Networks}

\author{Abdullah~Alhadlaq, Said~Kerrache, Hatim~Aboalsamh
\IEEEcompsocitemizethanks{\IEEEcompsocthanksitem King Saud University, College of Computer and Information Sciences, Riyadh, 11543, KSA.\protect\\
E-mail: 436106753@student.ksu.edu.sa}
}

\IEEEtitleabstractindextext{
\begin{abstract}
Recommender systems have become an essential tool for providers and users of online services and goods, especially with the increased use of the Internet to access information and purchase products and services. This work proposes a novel recommendation method based on complex networks generated by a similarity-popularity model to predict ones. We first construct a model of a network having users and items as nodes from observed ratings and then use it to predict unseen ratings. The prospect of producing accurate rating predictions using a similarity-popularity model with hidden metric spaces and dot-product similarity is explored. The proposed approach is implemented and experimentally compared against baseline and state-of-the-art recommendation methods on 21 datasets from various domains. The experimental results demonstrate that the proposed method produces accurate predictions and outperforms existing methods. We also show that the proposed approach produces superior results in low dimensions, proving its effectiveness for data visualization and exploration. 
\end{abstract}

}

\maketitle

\IEEEdisplaynontitleabstractindextext

\IEEEpeerreviewmaketitle

\IEEEraisesectionheading{\section{Introduction}\label{sec:introduction}}
A recommendation system (RS) is a software system that gathers users' actions and information as inputs to predict their future behavior by utilizing various techniques and then advises them with item recommendations that best fit their preferences \cite{Bobadilla2013Survey}. Items may refer to any product or service across various domains, including online dating, travel destinations, books, movies, and news.
Recommendation systems aim to lessen the information overload problem and reduce the enormous number of choices a user faces in online systems by highlighting only the most pertinent items \cite{Lu2015Survey}. 

Depending on the data used to generate recommendations, recommender systems can be broadly classified into two categories: collaborative filtering and content-based recommendation systems. In collaborative filtering (CF), the recommendations are based on the similarities between user preferences. The premise is that agreement in taste between users is sustained over extended periods, and past data can be used to predict future preferences \cite{handbook2015chapter2}. In content-based recommendation, on the other hand, the system recommends items with similar characteristics to items the user has positively rated. In other words, the recommendations are determined solely by items' features, properties, or descriptions.

Networks or graphs are powerful data representations naturally occurring in various real-world settings. Many applications have adopted graphs for their usefulness in providing insights into the intrinsic structure of the relationships within data \cite{goyal2018graph,cai2018comprehensive}. Network-based or graph-based RS models, where users and items are represented as nodes connected by weighted links that represent their similarity or relevancy, can achieve high accuracy while effectively handling data sparsity and limited coverage issues. 
Recently, many network-based approaches have been introduced into the field of RS, including networks based on physics theories such as complex networks \cite{huang2007analyzing, zhou2007bipartite, zanin2008complex}. Using a complex network to guide the recommendation process is a promising direction of research within the field \cite{boguna2010sustaining}. However, most of the work in this direction uses ad-hoc networks with no underlying model or established properties. 

In this work, we propose a novel recommendation method that uses a complex network generated by a similarity-popularity model to predict ratings. Users and items are represented as nodes in a graph possessing a complex network structure with an underlying hidden metric space model controlling the connection between users and items. We estimate the positions of users and items in a multidimensional Euclidean space from the observed user-items ratings. The coordinates of a user represent the user's preferences, and similarly, for items, the coordinates of an item represent its characteristics. The predicted ratings are then determined using the connection probabilities computed based on the similarity between nodes and their popularity.

Our main contributions are summarized as follows: 
\begin{itemize}
	\item We develop a novel similarity-popularity recommendation method that takes advantage of the latest developments in the field of complex networks.
	\item We adopt a complex network model, namely a similarity-popularity model, to direct the recommendation process. First, the model parameters are estimated from observed user-item ratings. The learned model is then used to predict unknown ratings while providing the ability to visualize and justify the results. 
	\item We conduct extensive experiments on several public datasets from various domains to evaluate the proposed method's effectiveness. The experiment results show that our approach produces accurate predictions and outperforms state-of-art methods.
\end{itemize}

The rest of this paper is organized as follows. Section \ref{sec:background} presents background on complex networks and their models and reviews the related work on network-based RS methods. Section \ref{sec:proposed} introduces the proposed method. In Section \ref{sec:results}, we conduct a detailed experimental analysis to evaluate the performance of the proposed method. Finally, we conclude the paper and highlight future research directions in Section \ref{sec:conclusion}.

\section{Background and Related Work} \label{sec:background}
This section covers essential concepts from the field of complex networks with emphasis on network models, in particular similarity-popularity models. We then give an overview of recent related research in the field of recommender systems.

\subsection{Complex networks}
The interest in studying complex networks arises from the fact that most real networks are naturally structured as complex networks. The nontrivial topological features of complex networks can be seen in virtually all types of networks, such as social, technological, and biological networks. For example, in real-life social networks, the interaction between people can be represented as a complex network. The nodes in this graph are the people, and the edges are the relationships among them \cite{costa2011complexsurvey}. Another natural network that adheres to the structure of complex networks is that of academic citations among papers, where nodes are the papers, and a directed edge indicates a citation \cite{newman2001collaboration}.
Complex networks also include technological networks that serve or deliver resources, such as the electric grid network or the network of roads and airports. One of the most studied examples of this type of network is the Internet which exhibits the structure of a scale-free network \cite{boguna2010sustaining,barabasi2000scale}. 
Food webs are also one of the most studied types of complex networks. In a food web, species are nodes, and directed edges between species are created when one species preys on the other \cite{williams2002food}. Recent studies show that trust networks  \cite{yuan2010small-worldness,trifunovic2010social,zhou2011emergence} in collaborative filtering systems also possess complex network characteristics, specifically those of scale-free networks \cite{cano2006music,yuan2013recommender}.

\subsubsection{Complex networks models}
\begin{figure*}
	\centering
	\includegraphics[width=0.3\linewidth]{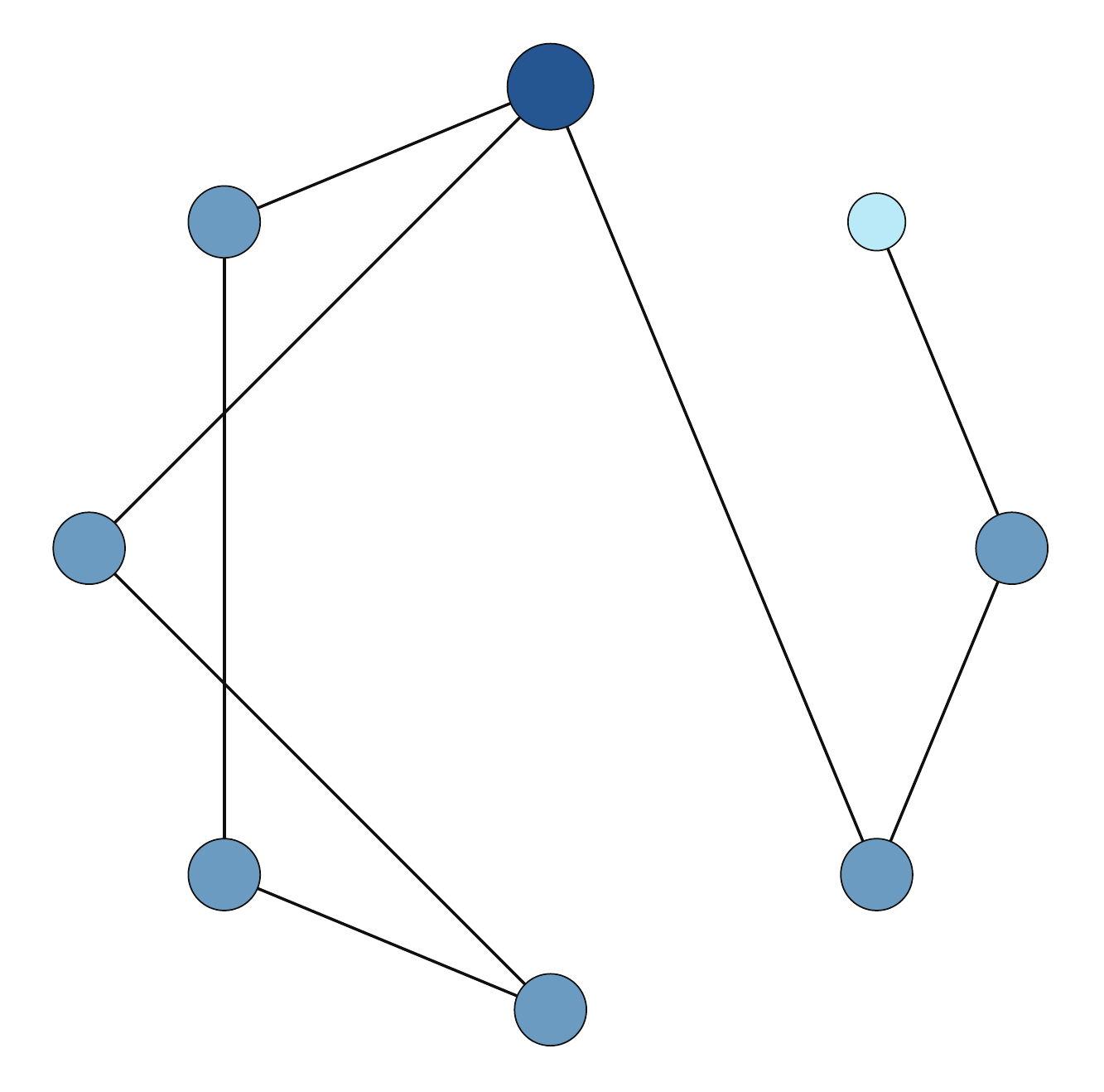}
	\quad
	\includegraphics[width=0.3\linewidth]{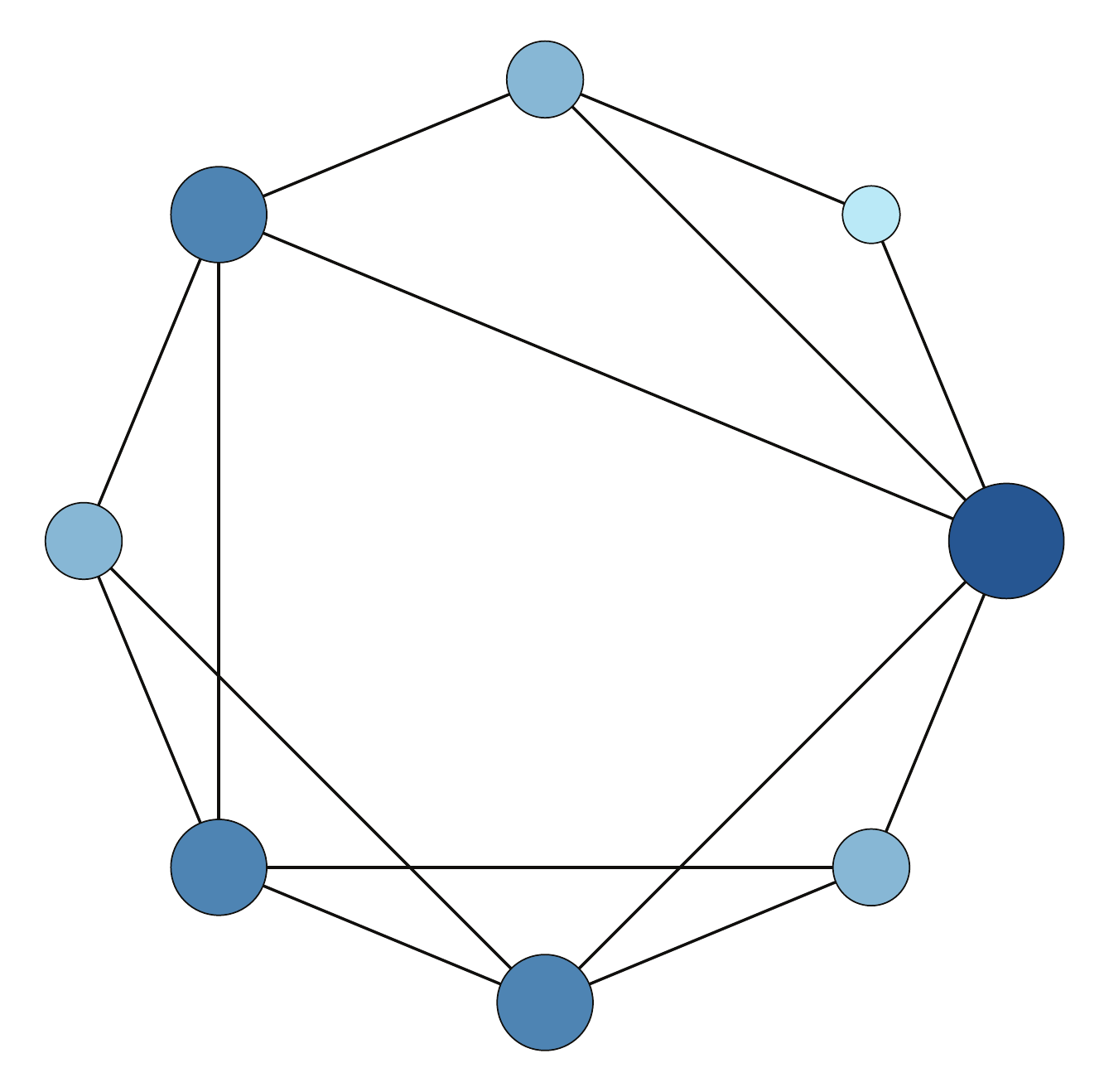}
	\quad
	\includegraphics[width=0.3\linewidth]{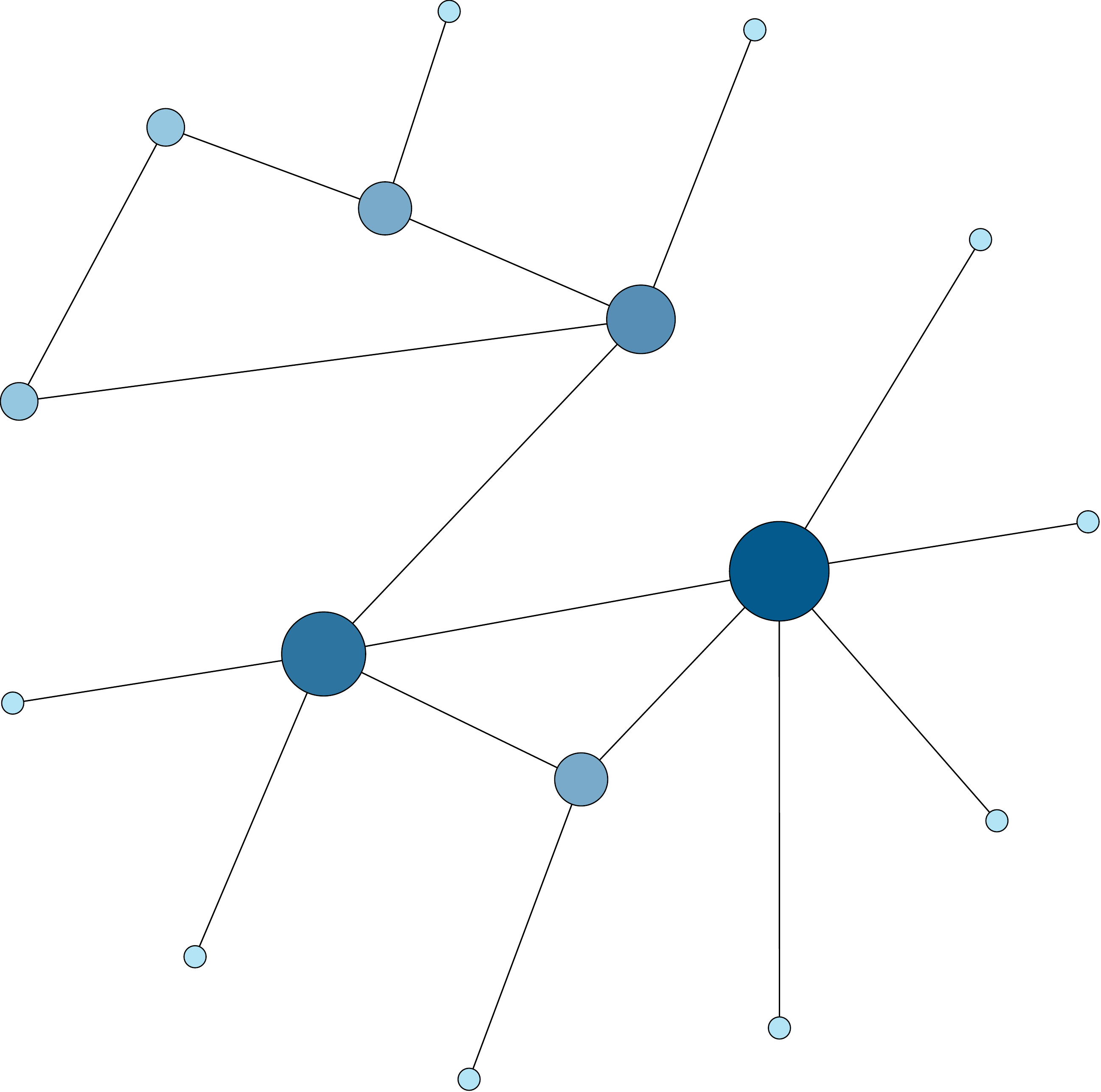}
	\caption{Illustrations of various complex network models (a) A random model. (b) A small-world model with short paths. (c) A scale-free model with short paths and highly connected nodes or hubs. The node's size and color reflect its degree.}
	\label{fig:models}
\end{figure*}

Initially, complex networks have been studied as part of the mathematical graph theory field that traditionally focused on regular graphs. In 1959 Paul Erdős and Alfréd Rényi proposed a random network model \cite{ERmodel1959} structured as a simple, straightforward design of a large-scale graph with no apparent design principles. Scientists were influenced afterward by their work and treated all real-life networks as if they were random \cite{barabasi2009decade,wang2003complex}. However, later studies showed that most networks are, in fact, more complex than random. For this, the field of network science witnessed a rebirth when Watts and Strogatz proposed small-world networks in 1998 \cite{WSmodel998}, followed by scale-free networks one year later by Barabási and Albert \cite{BAmodel1999}. Since then, complex networks have been considered graphs with nontrivial topological features that often occur in natural structures such as social or biological networks. Networks with characteristics that are not purely random or regular are considered complex. Figure \ref{fig:models} shows some examples of the mentioned models.  This section will discuss scale-free complex network models and their properties.

Scale-free networks gained considerable interest within the network science community upon publishing the Barabási-Albert model \cite{BAmodel1999}, which generates networks with scale-free properties. It is now considered one of the models that most reflect real-world natural network properties. The Barabási-Albert model focuses on two main features that the previous model lacked, the preferential attachment property and dynamic growth of the network. It also shares certain characteristics with the Erdős-Rényi random network model \cite{ERmodel1959} and Watts-Strogatz small-world model \cite{WSmodel998}. These include the small-world property and the tendency to cluster, which do occur in the Watts-Strogatz model but not in simple networks such as the Erdős-Rényi model.

Real-world networks contain highly connected nodes, or hubs, that have a much higher degree than the average degree of the network. The preferential attachment property states that new nodes are more likely to connect with these hubs, which describes the "rich get richer" phenomenon observed in the real world \cite{barabasi2009decade}. Hub nodes do not appear in random or small-world models; their existence affects the degree distribution and changes it from a Poisson distribution in random and small-world networks to a power-law distribution with a long tail.

\subsubsection{Similarity-popularity models}
The preferential attachment principle relies solely on nodes' popularity, represented by their degree, to construct the network. Similarity-popularity models are a class of complex network models extending this principle to include similarity between nodes as a factor contributing to link formation.
The hidden metric space model \cite{serrano2008HMS} is an example of a similarity-popularity model in which similarity between nodes is encoded using a hidden metric space underlying the network. The one-dimensional ring or circle is the simplest suggested model to construct such space \cite{boguna2009navigabilityHMS}. This technique generates a hidden metric space for any scale-free network by first positioning nodes uniformly randomly along the ring. Each node is then assigned its hidden degree $\kappa$ following a power-law distribution $P(\kappa)\sim \kappa^{-\gamma}$, where $\gamma > 2$. Finally, each pair of nodes is connected with probability $r(d; \kappa,\acute{\kappa})$:
\begin{equation}
	r(d;\kappa,\acute{\kappa})=r(d/d_c)=(1+d/d_c )^{-\alpha},
\end{equation}
where $\alpha>1$,  $d$ is the distance between the two nodes, and  $d(c)\sim\kappa\acute{\kappa}$ is the characteristic distance of the two nodes, with $\kappa,\acute{\kappa}$ being their hidden degrees.
According to this formula, the likelihood of a connection between any two nodes increases when the hidden degree $\kappa$ grows and decreases as their distance $d$ increases. The generated network possesses three characteristics observed in complex real-life networks: A pair of nodes with large degrees is more likely to become connected even if the distance $d$ is large. A pair of nodes with only one node having a high degree is connected if the distance $d$ is moderate. Final, a pair of unpopular nodes with low degrees is connected if and only if the distance $d$ between them is small.

Closely related to the hidden metric space model is the popularity similarity optimization model (PSO) \cite{muscoloni2018nonuniform}, which combines scale-free properties with hidden metric space properties to generate networks in a hyperbolic space where angular and radial coordinates represent nodes' positions. In PSO, the probability of connecting two nodes does not rely exclusively on popularity but on optimizing a balance between node popularity, abstracted by their radial coordinate and similarity, abstracted by the distance between their angular coordinates. As a result, PSO generates networks that possess a scale-free structure with degrees following power-law distribution and present many properties of real-world networks. 

Similarity-popularity models have been successfully applied to various tasks including, network modeling \cite{boguna2021network}, information routing \cite{boguna2009navigabilityHMS} and link prediction \cite{Papadopoulos2015NetworkMappingbyReplayingHyperbolicGrowth, kerrache16, kerrache20}. 

\subsection{Related work}
Using networks to guide the recommendation process has attracted increasing interest within the research community. The earliest work done in this direction is presented in \cite{aggarwal1999horting}, where users are represented by nodes in a directed graph and connected according to their similarity. Later on, many recommendation approaches adopted graph representations at their core \cite{Shi2014Survey}, either to model users and items as a bipartite graph, where users and items lie in separate groups with edges representing ratings, or as a single network where both users and items co-exist in the same space. 

More recently, authors in \cite{lee2015escaping} proposed a graph-based recommendation method by constructing a connected undirected graph to model their system. Their method (GraphRec) aims to find accurate yet novel items. In their work, only items are represented as nodes in the graph, and links between them indicate positive correlations. After constructing the graph, observed user ratings are used to filter the items. Positively correlated items ensure the relevancy of an item to the user preference, providing hence accurate recommendations. This idea is similar to our work. However, in our case, we position both items and users within the graph using a similarity-popularity model.

Authors in \cite{khoshneshin2010collaborative} proposed a collaborative filtering method that embeds both users and items in a Euclidean space, where the distance between nodes is inversely proportional to their ratings. Thus, the distance between users and items is shorter for items rated high. In their work, the training phase begins by constructing a model that finds the ideal location of each item and user in a high-dimensional Euclidean space based on observed ratings. Recommending items to a user is then done by searching for the nearest items in the space. The accuracy of their method was validated by experiments and showed improved results. Even though their method and the proposed method in our paper share similar fundamentals, the distances between items and users are not directly taken from the observed ratings in our case. Instead, we use an underlying model with additional factors to construct the graph. As a result, the generated network resembles real-world networks, including the existence of hubs and power-law degree distribution.

The authors construct a bipartite graph to represent items and users separately in another work \cite{zanin2008complex} focusing on applying analytical techniques used in complex networks to implement graph-based CF. The distances between nodes are calculated using cosine similarity. Once the graph is constructed, items are recommended to users following the preferential attachment principle, whereby popular items are assumed to be more likely to be chosen by a user. Even though the intuition behind popular items recommendation can be accepted as a solution to overcome the cold-start issue when user preferences are unknown, this approach has the obvious limitation of recommending popular items at the expense of relevant ones. Other related works that focused on using networks include guiding recommendation process by random graph modeling \cite{huang2007analyzing}, which confirms the feasibility of modeling users-items as random graphs. However, the characteristics of real-life networks do not match those of purely random graphs since real-world user-item interactions are more complex and contain a higher tendency to form clusters. Hence, it would be more beneficial to use a more realistic underlying model to better reflect the connections between users and items.  

Several approaches represent trust relationships as a graph, where links between users indicate their trust instead of their similarity. In trust-based networks, users can trust each other either explicitly \cite{walter2008trust} or implicitly \cite{azadjalal2017trust} through their ratings or other observed actions. Once the trust between users has been established, it can be used to guide the recommendations process as in similarity-based CF. In \cite{walter2008trust}, the authors propose a recommendation system that takes advantage of trust between users in an online social network. They obtained favorable results by proposing a model that uses the predefined relationships between users in the network to reach and filter information and knowledge. Their model consists of a bipartite graph of users and object nodes representing recommended items, and the weights of the connection between users and objects are ratings. In our case, instead of using a predefined neighborhood produced from the friendship network, the network is constructed based on previous ratings, similarities, and popularities of items without an additional source of information.

The experimental results obtained by the authors in \cite{yuan2010small-worldness} confirm the viability of using a scale-free network for recommendation as they experimentally show that trust networks possess the small-world property in which the network diameter is significantly small compared to the number of nodes. They use this property to estimate the trust between users based on the distance that separates them in the network. The computed trust value is then used to weigh the recommendations obtained from different users. The trust network approach also overcomes the cold-start and data sparsity issues. The authors in \cite{azadjalal2017trust} argue that most e-commerce RSs users do not have explicitly trusted users, either because the functionality to establish trust between users explicitly is unavailable or due to cold-start and data sparsity issues. Thus, they proposed constructing a weighted directed trust network for each user using both explicit and implicit trust. The network consists of nodes representing users' neighbors and only those are used to compute the recommendations. 

Using trust in recommender systems increases the resistance to malicious attacks. However, the trust network's primary purpose is to overcome the cold-start issue. New users can find trusted users even if the system cannot find similar users as their preferences are still unknown. However, the main drawback of trust-based recommendation is the need for additional information that is not always present nor easy to elicit from data. In this paper, we build a network from ratings only, which broadens the applicability of our proposed approach.

\section{The Proposed Method}
\label{sec:proposed}
This paper proposes a recommendation approach based on a similarity-popularity model, where users and items are modeled as nodes in a graph having a scale-free network structure. Before delving into the details of the approach, however, we start by introducing notation. We consider a recommendation system consisting of $n$ users and $m$ items and denote by $r_{ij}$ the rating given by user $i$ to the item $j$ and by $R$ the set of all ratings. We denote by $\bar{r}_i$ the average rating for user $i$ and by $\bar{r}_j$ the average rating for item $j$. The minimum and maximum ratings are denoted by $r_{min}$ and $r_{max}$, respectively. We will often need to transform the ratings into probabilities. For this, we define the set $\tilde{R}$ of scaled ratings $\tilde{r}_{ij}$ obtained from raw ratings by applying the function $\phi$ defined as:
\begin{equation}
	\label{eq:phi}
	\tilde{r}_{ij} \equiv \phi(r_{ij}) = \dfrac{r_{ij} - r_{min}}{r_{max} - r_{min}} \left(p_{max}-p_{min} \right) + p_{min},
\end{equation}
where $p_{min}$ and $p_{max}$ are the minimum and maximum probabilities, respectively. The function $\phi$ maps the raw ratings to the closed interval $[p_{min}, p_{max}]$. To avoid boundary issues, we exclude the two extremities 0 and 1 when selecting $p_{min}$ and $p_{max}$ so that we have $0< p_{min}, p_{max} <1$. We also define the translation $\varphi$ as follows:
\begin{equation}
	\label{eq:varphi}
	\varphi(r) = r - r_{min}+1.
\end{equation}

In the proposed approach, each user $i$ is assigned a point $x^i$ in a $D$-dimensional Euclidean space:
\begin{equation}
	x^i=\left(x^i_1, x^i_2, \ldots, x^i_D\right)^T , i = 1,\ldots, n.
\end{equation}
The users' coordinates reflect their preferences concerning the nature of the recommended items. Hence, each hidden dimension represents a hidden feature relevant to the recommendation or a combination of several such features. For example, in a movie recommendation system, relevant features may be the length of the movie the user usually watches or how much action, drama, or comedy the user prefers. Likewise, each item is assigned a position $y^j$ in the same  $D$-dimensional Euclidean space.
\begin{equation}
	y^j=\left(y^j_1, y^j_2, \ldots, y^j_D\right)^T , j = 1,\ldots, m.
\end{equation}
Likewise, the position of an item represents its description or list of features. 

The rating $r_{ij}$ given by user $i$ to item $j$ is assumed to be proportional to the probability of connection between the two corresponding nodes, that is,  $r_{ij} \propto {p}_{ij}$, where ${p}_{ij}$ is prescribed by a similarity-popularity model. The first model we propose is a similarity-popularity model based on a hidden metric space model inspired by \cite{serrano2008HMS} and will henceforth be referred to as SPHM1. The connection probability under SPHM1 is defined as:
\begin{equation} 
	\label{eq:SPHM1}
	{p}^{SPHM1}_{ij}={\left(1+\dfrac{d^2(x^i,y^j)}{\sqrt{\kappa_i^{SPHM1}\kappa_j^{SPHM1}}}\right)}^{-\alpha}, 
\end{equation}
where $\kappa_i^{SPHM1}$ and $\kappa_j^{SPHM1}$ are the expected degrees of $i$ and $j$ respectively, and $d(x^i,y^j)$ is the hidden Euclidean distance between them. Hence, the connection probability decreases with the squared hidden distance $d^2(x^i,y^j)$ (dissimilarity)  and increases with the popularity $\kappa_i^{SPHM1}$ and $\kappa_j^{SPHM1}$. The rationale is that users tend to favor items that fit their preferences, that is, items that are close to them in the Euclidean space. However, items with a high degree can be liked by users even if they are considerably different from what they usually prefer. Also, some items having a small or average degree may be liked by faraway users having a large degree. 

\begin{figure*}[!t]
	\centering
	\includegraphics[width=0.32\textwidth,  height=0.32\textwidth]{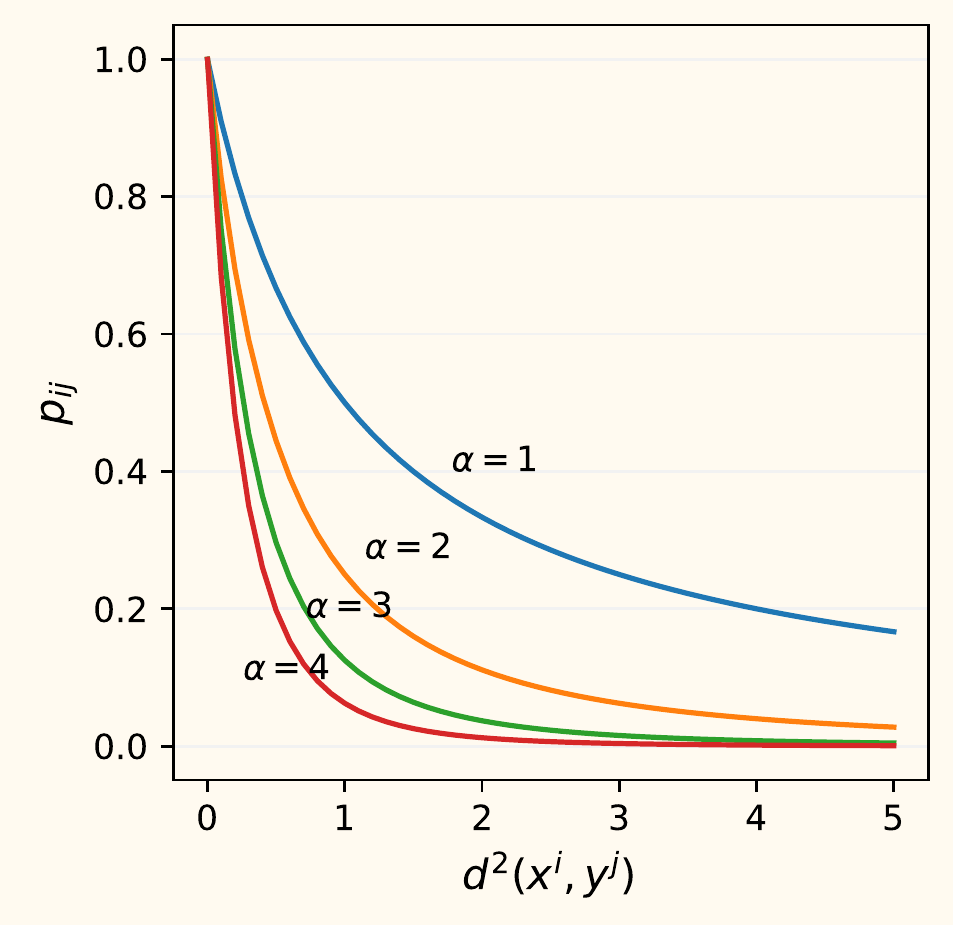}
	\includegraphics[width=0.32\textwidth,  height=0.32\textwidth]{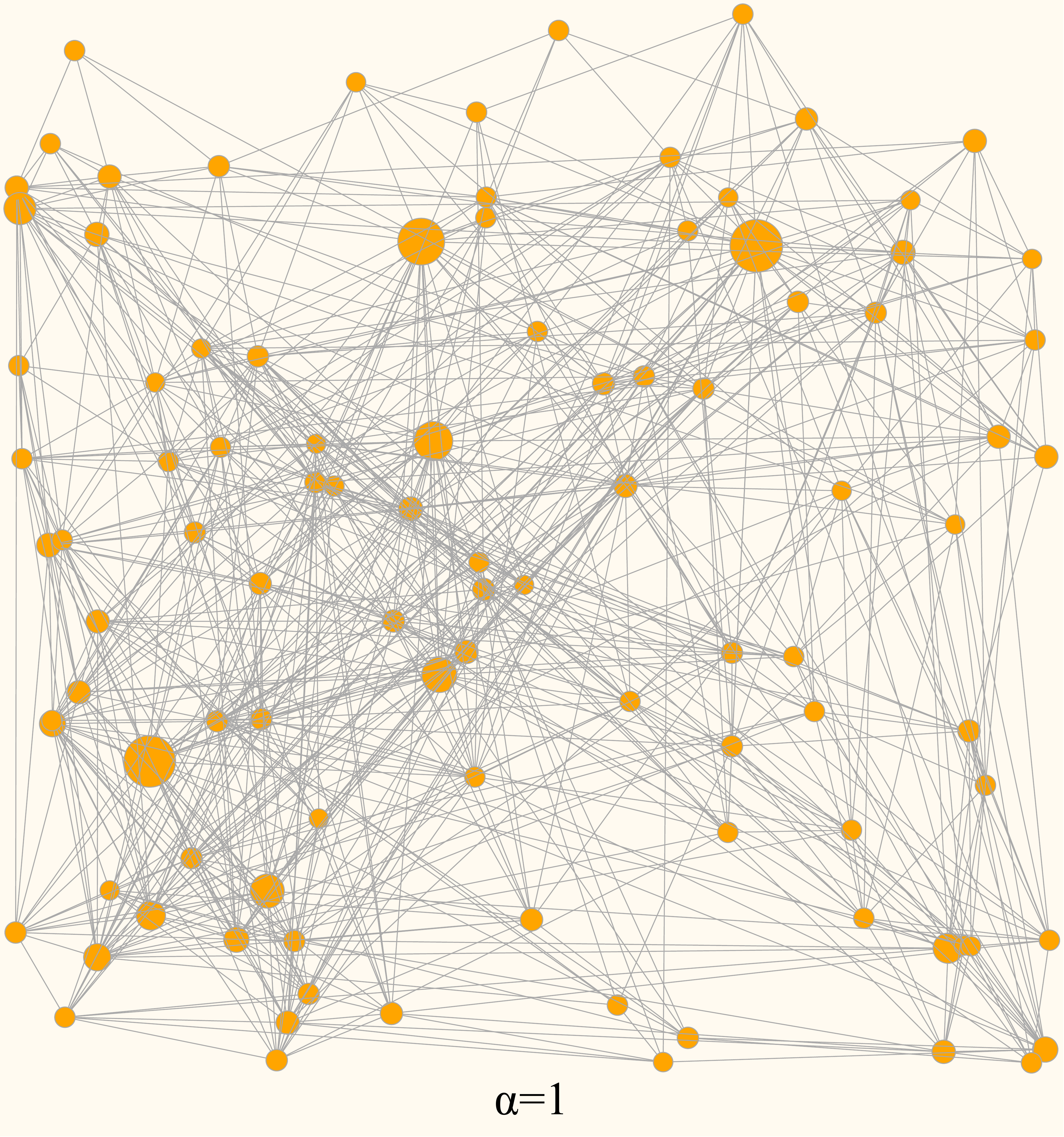}
	\includegraphics[width=0.32\textwidth,  height=0.32\textwidth]{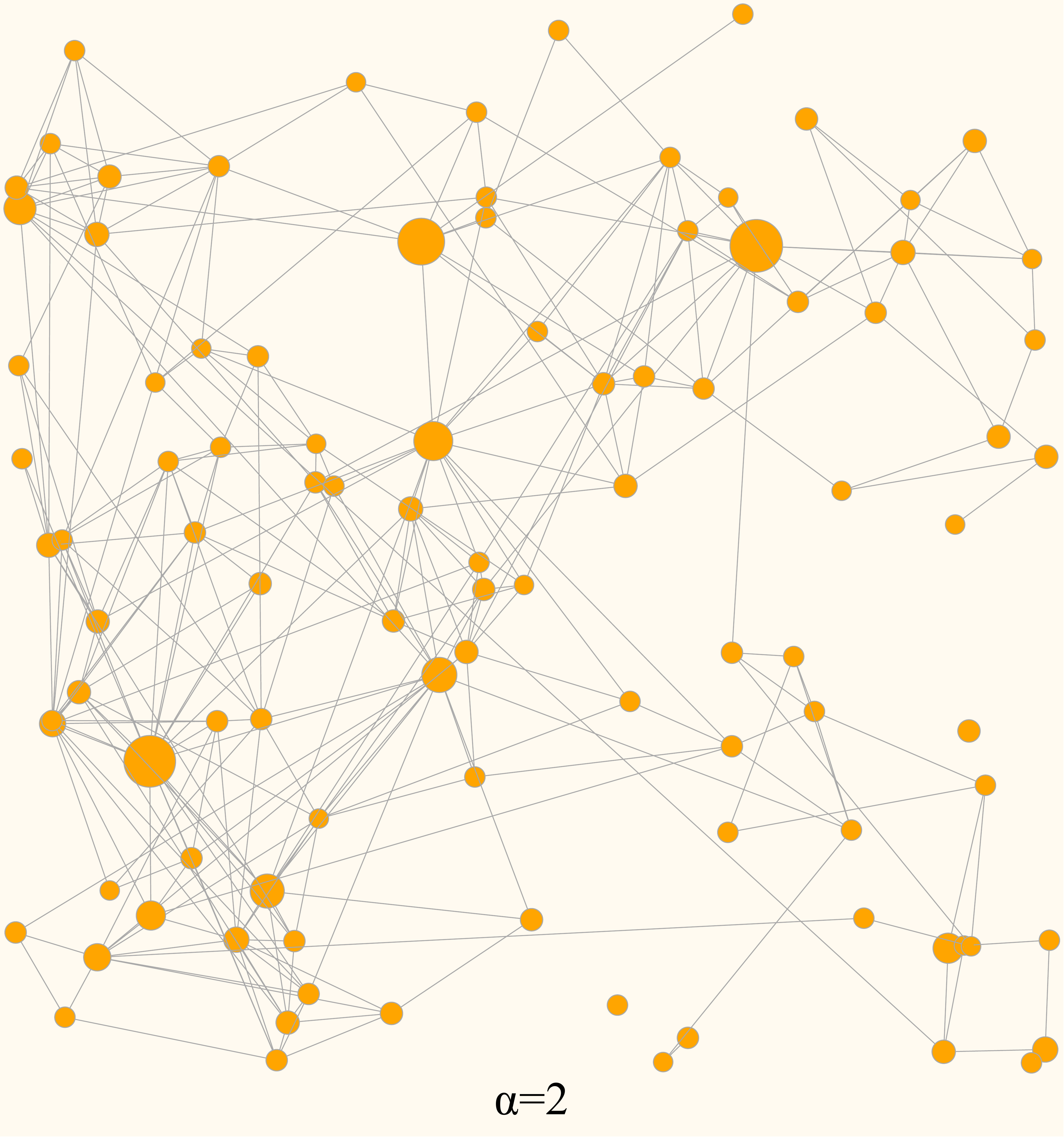}
	
	\includegraphics[width=0.32\textwidth,  height=0.32\textwidth]{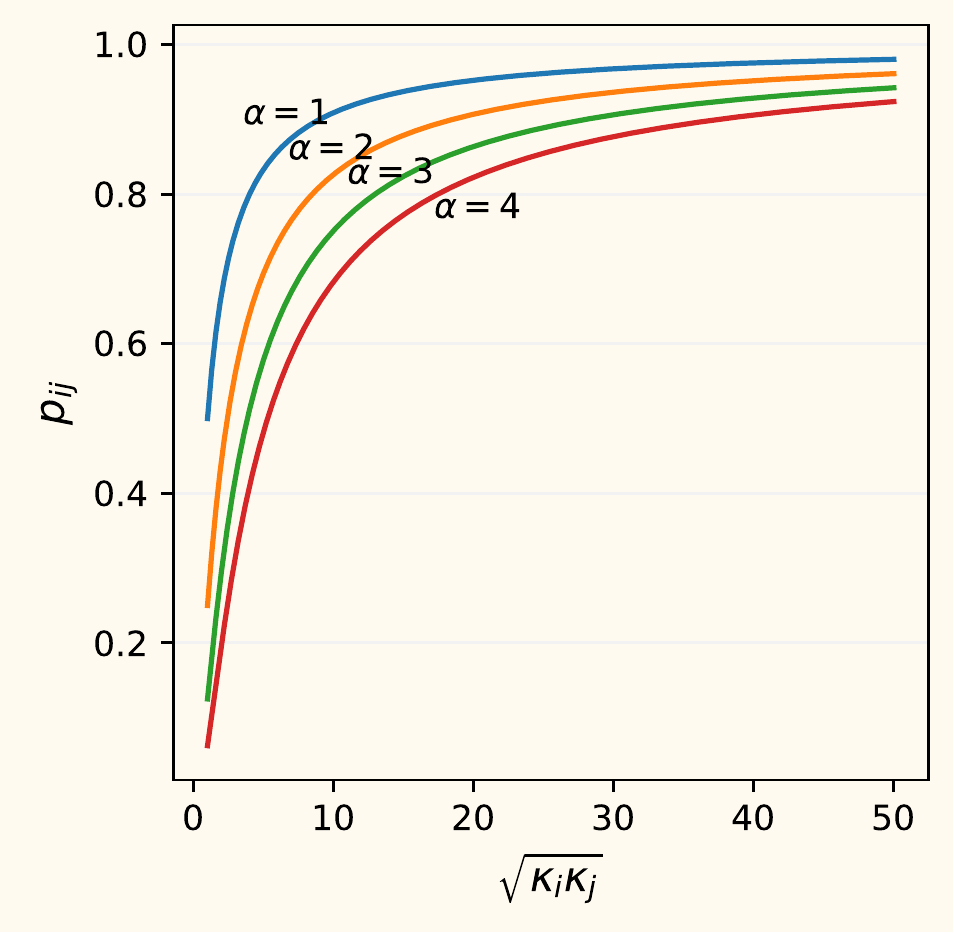}
	\includegraphics[width=0.32\textwidth,  height=0.32\textwidth]{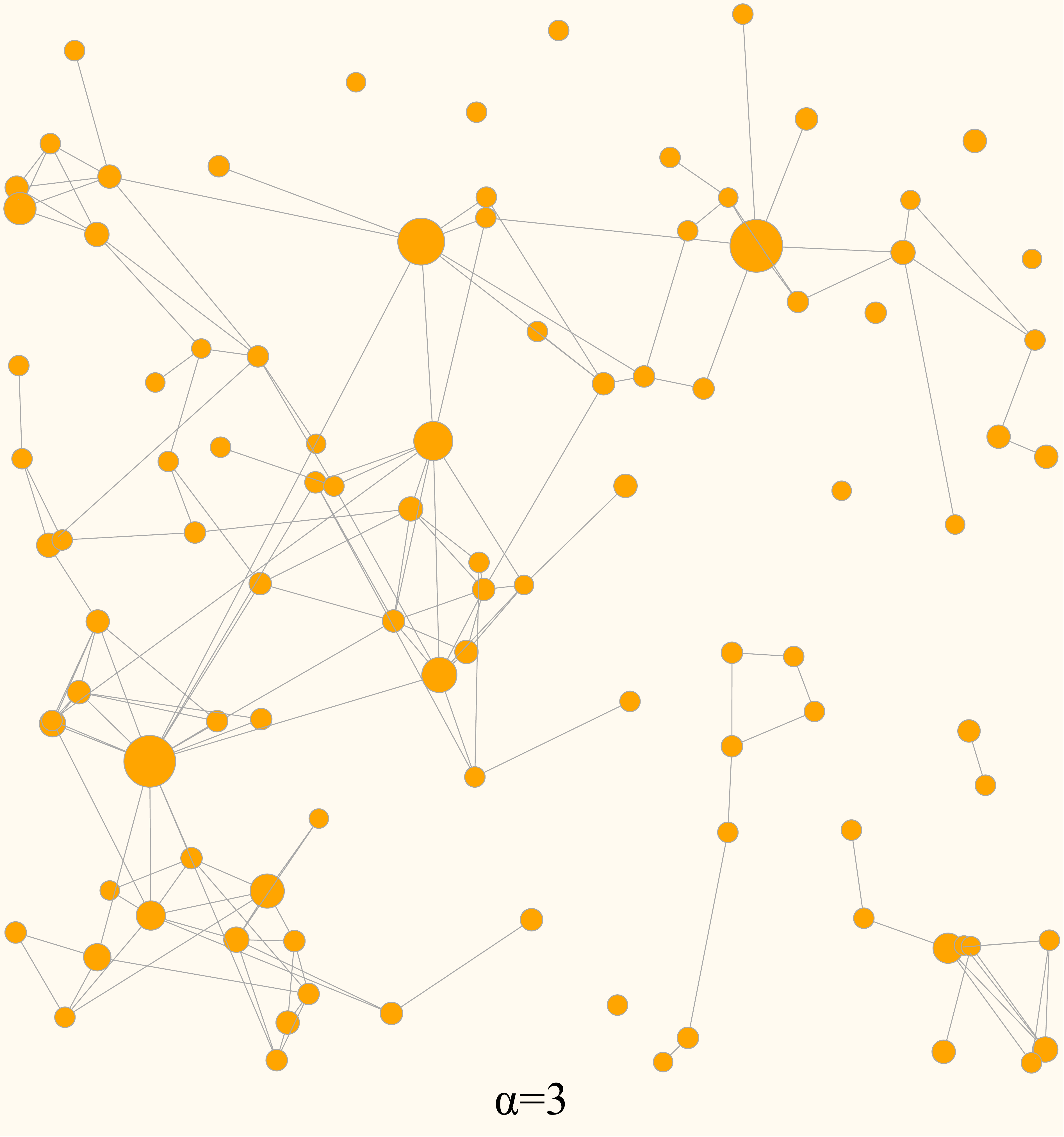}
	\includegraphics[width=0.32\textwidth,  height=0.32\textwidth]{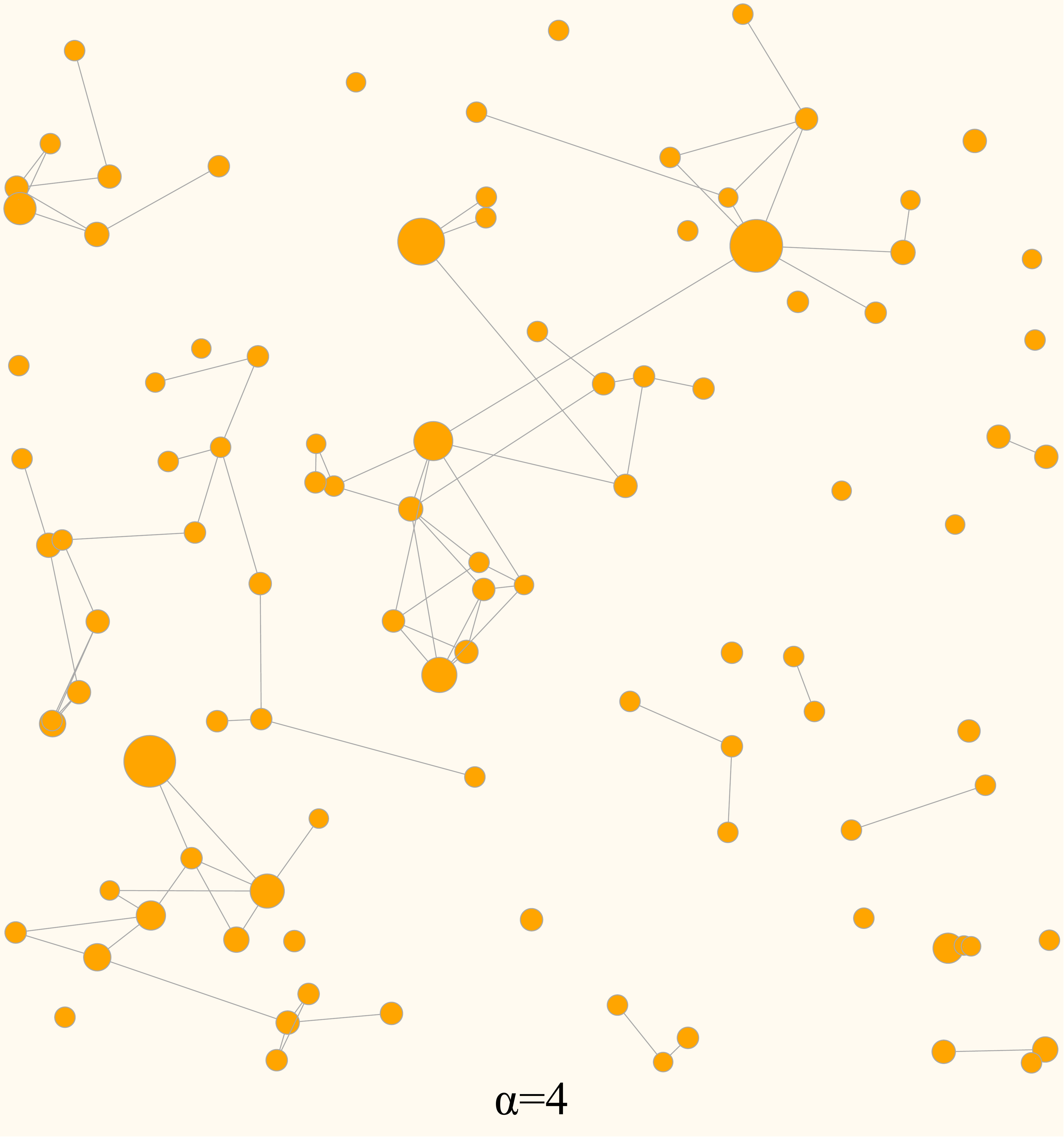}
	\caption{Effect of the parameter $\alpha$ on the network topology. The two line-plots show the evolution of the connection probability $p_{ij}$ as a function of the squared distance with fixed popularity (upper left) and popularity with fixed distance (bottom left) for different values of $\alpha$. The graph plots show one sample network resulting from each value of $\alpha$.}
	\label{fig:alpha}
\end{figure*}

The parameter $\alpha$ controls the network clustering and the effect of the distance and popularity on the connection probability. As illustrated in Figure \ref{fig:alpha}, larger values of $\alpha$ cause a faster decay in probability as the distance grows and a slower increase in probability as popularity increases. Consequently, with everything else fixed, low values of $\alpha$ result in denser graphs. In contrast, larger values of $\alpha$ result in sparser, more clustered graphs with mostly short-range connections and long-range connections only occurring between highly popular nodes.

The second similarity-popularity model (SPHM2) is deduced from the first model by defaulting the value of $\alpha$ to $1$, which results in a simpler model which requires less effort for parameter tuning and validation: 
\begin{equation} 
	\label{eq:SPHM2}
	p^{SPHM2}_{ij}=\left(1+{\dfrac{d^2(x^i,y^j)}{\sqrt{\kappa_i^{SPHM2}\kappa_j^{SPHM2}}}}\right)^{-1}.
\end{equation}

The third model uses the dot product to encode the similarity between nodes instead of the Euclidean distance. This results in a gradient that is easier to compute and thus faster optimization:
\begin{equation} 
	\label{eq:SPDP}
	p^{SPDP}_{ij}= \sqrt{\kappa_i^{SPDP}\kappa_j^{SPDP}} \exp\left({x^i \cdot y^j}\right).
\end{equation}
Unlike the two previous models, $p^{SPDP}$ is not restricted to the interval $[0, 1]$ and hence does not qualify as a valid connection probability. We can instead think about it as a score assigned to node couples, where higher scores indicate a higher likelihood of connection, which is a common practice in the link prediction literature \cite{Lu2011LinkPredictioninComplexNetworks}.

\begin{figure*}	
	\centering
	\includegraphics[scale=1.3]{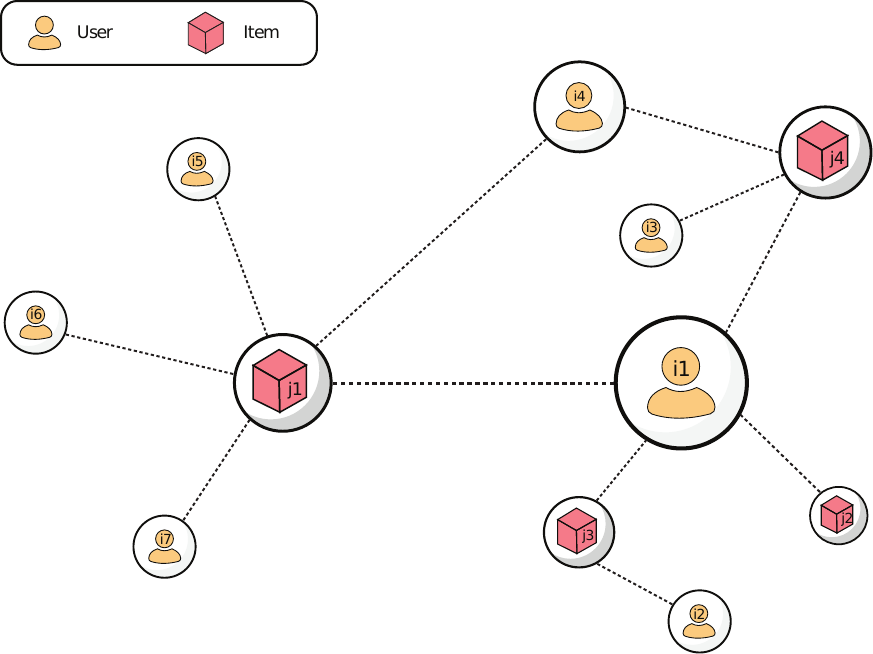}
	\caption{The proposed approach models the system as a complex network where users and items connect according to similarity and popularity. The dissimilarity between nodes is encoded by the distance separating them, whereas popularity is an intrinsic property of the node represented here by its size.}
	\label{fig:proposedModelGraph}
\end{figure*}

As illustrated in Figure \ref{fig:proposedModelGraph}, users and items in the proposed approach co-exist in a graph embedded in a $D$-dimensional space. They are connected based on two factors, their similarity (or distance) and popularity (or degree). For instance, item $j_1$ is popular, which increases its probability of connecting to distant dissimilar users, especially if they have large degrees, such as user $i_1$. For the same reason, the latter is highly likely to connect to many items. On the other hand, low-degree items such as $j_2$ have less probability of connecting to users. Thus, it will not be able to connect to nearby users with a low degree such as $i_2$. This phenomenon can be observed in real life, where some movies, for instance, have a broad audience and are enjoyed by people who do not usually appreciate that type of movies. Similarly, some moviegoers have a taste for a wide range of movie genres and styles.

Figure \ref{fig:proposedModelGraph} also hints at the possibility of using the model for visualization, which requires training the model in two or three dimensions. Such a low dimensionality, however,  can negatively impact recommendation accuracy. One can overcome this issue by training the model in a high-dimensional space, then picking the desired neighborhood of users and items and their distances, and finally, rebuilding the space in two or three dimensions using any graph embedding technique. 

To learn the proposed models from  data, we first set the users' and items' degrees. For the two first models, SPHM1 and SPHM2, we set the degrees as follows:
\begin{align}
	\label{eq:kappa-hm}
	\kappa_i^{SPHM1} = \kappa_i^{SPHM2} &= \varphi(\bar{r}_i), \quad i =1,\ldots, n\\
	\kappa_j^{SPHM1} = \kappa_j^{SPHM2} &= \varphi(\bar{r}_j), \quad j =1,\ldots, m.
\end{align}
This step guarantees that all degrees are greater than or equal to one. For the SPDP model, we transform the average ratings using the function $\phi$ instead:
\begin{align}
	\label{eq:kappa-dp}
	\kappa_i^{SPDP} &= \phi(\bar{r}_i), \quad i =1,\ldots, n\\
	\kappa_j^{SPDP} &= \phi(\bar{r}_j), \quad j =1,\ldots, m.
\end{align}
Once the node degrees are fixed, computing the distances between nodes amounts to finding the coordinates of each user and each item so that the predicted ratings are as similar as possible to the actual ratings. This problem can be formulated as that of minimizing the following objective function:
\begin{multline}  
	\label{eq:L2}
	J_{L_2}(x^1, \ldots, x^n, y^1, \ldots, y^m) = \\ \sum_{\tilde{r}_{ij}\in \tilde{R} } \left({p}_{ij} - \tilde{r}_{ij}\right)^2 + \lambda  \left( \sum_{i=1}^{n} \|{x^i}\|^2_2 + \sum_{j=1}^{m} \|{y^j}\|^2_2\right),
\end{multline}
where ${p}_{ij}$ is is the predicted probability or score computed by one the proposed models (Eq. \eqref{eq:SPHM1}, \eqref{eq:SPHM2} or \eqref{eq:SPDP}), $\lambda$ is the regularization coefficient, and $\|\cdot\|_2$ stands for the $L_2$ norm. The hyper-parameter $\lambda$ is chosen from a pre-defined set of values during the validation phase. Instead of using the squared error and $L_2$ normalization, it is possible to use the absolute error with $L_1$ normalization:
\begin{multline}  
	\label{eq:L1}
	J_{L_1}(x^1, \ldots, x^n, y^1, \ldots, y^m) = \\ \sum_{\tilde{r}_{ij}\in \tilde{R} } \left| {p}_{ij} - \tilde{r}_{ij} \right| + \lambda  \left( \sum_{i=1}^{n} \|{x^i}\|_1 + \sum_{j=1}^{m} \|{y^j}\|_1\right). 
\end{multline}

The problem of minimizing the objective functions \eqref{eq:L2} and \eqref{eq:L1} falls into the category of nonlinear optimization problems. Since the objective function is not convex, only local minima can be computed in general. Several algorithms for local optimization exist that can be used to solve this minimization problem, such as simple gradient descent or conjugate gradient. For completeness, we include the gradients of all combinations of the two cost functions $J_{L_2}$ and $J_{L_1}$ with the three proposed models in Table \ref{tab:gradient}.

\begin{table*}[!t]
	\caption{Gradients of all combination of cost functions and proposed models. The function $\sign$ returns 1 if its argument is positive, -1 if  negative, and 0 if null.}
	\label{tab:gradient}
	\centering
	\begin{tabular}{lll}
		\toprule
		Model & Cost & Gradient \\ \midrule
		\multirow{4}{*}{SPHM1}   & \multirow{2}{*}{$J_{L_2}$}  & $
		\dfrac{\partial J_{L_2}}{\partial x^i_k} =\sum_{\left\{j|\tilde{r}_{ij}\in \tilde{R} \right\} } -\dfrac{4\alpha}{\sqrt{\phi(\bar{r}_i) \phi(\bar{r}_j)}} \left(p_{ij} - \tilde{r}_{ij}\right) \left( 1+ \dfrac{d^2(x^i, y^j)}{\sqrt{\phi(\bar{r}_i) \phi(\bar{r}_j)}}\right)^{-\alpha-1} \left(x^i_k-y^j_k \right )+ 2 \lambda x^i_k
		$ \\ 
		&  & $
		\dfrac{\partial J_{L_2}}{\partial y^j_k} =\sum_{\left\{i|\tilde{r}_{ij}\in \tilde{R} \right\} } -\dfrac{4\alpha}{\sqrt{\phi(\bar{r}_i) \phi(\bar{r}_j)}} \left(p_{ij} - \tilde{r}_{ij}\right) \left( 1+ \dfrac{d^2(x^i, y^j)}{\sqrt{\phi(\bar{r}_i) \phi(\bar{r}_j)}}\right)^{-\alpha-1} \left(y^j_k-x^i_k \right )+ 2 \lambda y^j_k
		$ \\ \cmidrule{3-3}
		& \multirow{2}{*}{$J_{L_1}$}  & $
		\dfrac{\partial J_{L_1}}{\partial x^i_k} =\sum_{\left\{j|\tilde{r}_{ij}\in \tilde{R} \right\} } -\dfrac{2\alpha}{\sqrt{\phi(\bar{r}_i) \phi(\bar{r}_j)}} \sign\left(p_{ij} - \tilde{r}_{ij}\right) \left( 1+ \dfrac{d^2(x^i, y^j)}{\sqrt{\phi(\bar{r}_i) \phi(\bar{r}_j)}}\right)^{-\alpha-1} \left(x^i_k-y^j_k \right )+ \lambda \sign(x^i_k)
		$ \\
		&   & $
		\dfrac{\partial J_{L_1}}{\partial y^j_k} =\sum_{\left\{i|\tilde{r}_{ij}\in \tilde{R} \right\} } -\dfrac{2\alpha}{\sqrt{\phi(\bar{r}_i) \phi(\bar{r}_j)}} \sign\left(p_{ij} - \tilde{r}_{ij}\right) \left( 1+ \dfrac{d^2(x^i, y^j)}{\sqrt{\phi(\bar{r}_i) \phi(\bar{r}_j)}}\right)^{-\alpha-1} \left(y^j_k-x^i_k \right )+ \lambda \sign(y^j_k)
		$ \\ \midrule
		\multirow{4}{*}{SPHM2}   & \multirow{2}{*}{$J_{L_2}$}  & $
		\dfrac{\partial J_{L_2}}{\partial x^i_k} =\sum_{\left\{j|\tilde{r}_{ij}\in \tilde{R} \right\} } -\dfrac{4}{\sqrt{\phi(\bar{r}_i) \phi(\bar{r}_j)}} \left(p_{ij} - \tilde{r}_{ij}\right) p_{ij}^{2} \left(x^i_k-y^j_k \right )+ 2 \lambda x^i_k
		$ \\
		&  & $
		\dfrac{\partial J_{L_2}}{\partial y^j_k} =\sum_{\left\{i|\tilde{r}_{ij}\in \tilde{R} \right\} } -\dfrac{4}{\sqrt{\phi(\bar{r}_i) \phi(\bar{r}_j)}} \left(p_{ij} - \tilde{r}_{ij}\right) p_{ij}^{2} \left(y^j_k-x^i_k \right )+ 2 \lambda y^j_k
		$ \\ \cmidrule{3-3}
		& \multirow{2}{*}{$J_{L_1}$}  & $
		\dfrac{\partial J_{L_1}}{\partial x^i_k} =\sum_{\left\{j|\tilde{r}_{ij}\in \tilde{R} \right\} } -\dfrac{2}{\sqrt{\phi(\bar{r}_i) \phi(\bar{r}_j)}} \sign\left(p_{ij} - \tilde{r}_{ij}\right) p_{ij}^{2} \left(x^i_k-y^j_k \right )+ \lambda \sign(x^i_k)
		$ \\
		&   & $
		\dfrac{\partial J_{L_1}}{\partial y^j_k} =\sum_{\left\{i|\tilde{r}_{ij}\in \tilde{R} \right\} } -\dfrac{2}{\sqrt{\phi(\bar{r}_i) \phi(\bar{r}_j)}} \sign\left(p_{ij} - \tilde{r}_{ij}\right) p_{ij}^{2} \left(y^j_k-x^i_k \right )+ \lambda \sign(y^j_k)
		$ \\ \midrule
		\multirow{4}{*}{SPDP}   & \multirow{2}{*}{$J_{L_2}$}  & $
		\dfrac{\partial J_{L_2}}{\partial x^i_k} =\sum_{\left\{j|\tilde{r}_{ij}\in \tilde{R} \right\} } 2 p_{ij} \left(p_{ij} - \tilde{r}_{ij}\right)  y^j_k + 2 \lambda x^i_k
		$ \\
		&  & $
		\dfrac{\partial J_{L_2}}{\partial y^j_k} =\sum_{\left\{i|\tilde{r}_{ij}\in \tilde{R} \right\} } 2 p_{ij} \left(p_{ij} - \tilde{r}_{ij}\right) x^i_k + 2 \lambda y^j_k
		$ \\ \cmidrule{3-3}
		& \multirow{2}{*}{$J_{L_1}$}  & $
		\dfrac{\partial J_{L_1}}{\partial x^i_k} =\sum_{\left\{j|\tilde{r}_{ij}\in \tilde{R} \right\} }  \sign\left(p_{ij} - \tilde{r}_{ij}\right) p_{ij} y^j_k + \lambda \sign(x^i_k)
		$ \\
		&   & $
		\dfrac{\partial J_{L_1}}{\partial y^j_k} =\sum_{\left\{i|\tilde{r}_{ij}\in \tilde{R} \right\} }  \sign\left(p_{ij} - \tilde{r}_{ij}\right) p_{ij} x^i_k + \lambda \sign(y^j_k)
		$ \\
		\bottomrule
	\end{tabular}
\end{table*}

In summary, the proposed models' parameters fall into three categories. The dimension $D$, the constant $\alpha$, and the regularization coefficient $\lambda$ are hyper-parameters determined during the validation phase using a simple grid search. The users' and items' degrees $\kappa_i$, $\kappa_j$ are computed directly from the observed ratings. Finally, the positions of users and items, $x^i$, $y^j$, are found by optimizing the objective functions $J_{L_2}$ and $J_{L_1}$.  The last step represents the bulk of the computational effort required to build the models.

\section{Experimental Evaluation} 
\label{sec:results}
In this section, we report the results of an extensive experimental analysis of the proposed method. We start by presenting the data used for the evaluation, the performance criteria, the competing methods, and the experimental setup. We then present and discuss the obtained results.

\subsection{Data}
The proposed method is compared to baseline and state-of-the-art methods found in the literature on 21 publicly available datasets. The datasets are of various domains and sizes and are widely used in the literature to evaluate recommender systems. Datasets were pre-processed to retain only users having at least five ratings. Table \ref{table:datasets} shows the datasets' names, description, and basic statistics.

\begin{table*}[!t]
	\caption{Datasets description and statistics.}
	\label{table:datasets}
	\centering
	\begin{tabular}{llrrr}
		\toprule
		Dataset                                  & Description            & Ratings   & Users  & Items   \\ \midrule
		AmazonAA   \cite{wan2016modeling}        & Apps   for Android     & 835,648   & 94,945 & 44,675  \\
		AmazonAuto   \cite{wan2016modeling}      & Automotive             & 284,101   & 34,892 & 129,108 \\
		AmazonBaby   \cite{wan2016modeling}      & Baby                   & 232,749   & 27,655 & 32,553  \\
		AmazonDM   \cite{wan2016modeling}        & Digital   Music        & 238,692   & 22,878 & 115,082 \\
		AmazonGGF  \cite{wan2016modeling} & Grocery   and Gourmet Food   & 329,057 & 32,228 & 84,357  \\
		AmazonHPC  \cite{wan2016modeling} & Health   and Personal Care   & 625,554 & 70,498 & 122,090 \\
		AmazonIV   \cite{wan2016modeling}        & Amazon   Instant Video & 63,836    & 8,072  & 11,830  \\
		AmazonMI   \cite{wan2016modeling}        & Musical   Instruments  & 87,791    & 10,057 & 36,099  \\
		AmazonPLG  \cite{wan2016modeling} & Patio,   Lawn and Garden     & 114,504 & 15,114 & 37,983  \\
		AmazonTHI  \cite{wan2016modeling} & Tools   and Home Improvement & 390,454 & 45,791 & 112,855 \\
		AmazonVG   \cite{wan2016modeling}        & Video   Games          & 300,003   & 31,027 & 33,899  \\
		Anime      \cite{Anime}                  & Anime                  & 6,318,424 & 60,970 & 9,927   \\
		Book-Crossing\cite{ziegler2005improving} & Books                  & 1,028,948 & 22,816 & 319,198 \\
		CiaoDVD    \cite{guo2014etaf}            & Movies                 & 49,382    & 2,609  & 14,312  \\
		Epinions   \cite{massa2008trustlet}      & General goods          & 631,192   & 23,253 & 137,289 \\
		FilmTrust  \cite{guo2013novel}           & Movies                 & 34,886    & 1,227  & 2,059   \\
		Food.com   \cite{majumder2019generating} & Food recipes           & 692,791   & 16,440 & 175,941 \\
		ML100K     \cite{Maxwell2015MovieLens}   & Movies                 & 100,000   & 943    & 1,682   \\
		ML1M       \cite{Maxwell2015MovieLens}   & Movies                 & 1,000,209 & 6,040  & 3,706   \\
		YahooMovies\cite{yahoo}                  & Movies                 & 221,364   & 7,642  & 11,916  \\
		YahooMusic   \cite{yahoo}                & Music                  & 365,704   & 15,400 & 1,000   \\ \bottomrule
	\end{tabular}
\end{table*}

The experiments use 5-fold cross-validation to maximize the results' accuracy. The data is randomly shuffled and split into five equal batches. Each batch is used once as a test set while the rest is used for training, resulting in 80\%-20\% splits for training and testing, respectively. The training set is further divided into a (proper) training set containing 90\% of the initial training samples and a validation test containing the remaining 10\% of the samples. The validation set is used to tune the model hyperparameters using grid search. Once the best parameters are found, the model is retrained on the whole training set before testing. Finally, the average performance over all five folds is reported.

\subsection{Evaluation criteria}
Several metrics have been proposed for evaluating recommendation systems. These metrics measure the accuracy of the predicted ratings provided by the recommendation method by comparing them to ground truth ratings. The model is used to predict a previously unseen rating by generating an estimate of it denoted by $\hat{r}_{ij}$, whereas $r_{ij}$ denotes the actual rating from the ground truth data. The most used rating prediction accuracy metrics are the Root Mean Squared Error (RMSE) and the Mean Absolute Error (MAE) \cite{portugal2017use}. These metrics allow the evaluation of the quality of numeric predictions. The MAE takes the sum of differences between the actual and predicted ratings and divides it by the number of ratings considered:
\begin{equation}
	MAE = \frac{1}{|R^{Test}|} \sum_{r_{ij} \in R^{Test}} | \hat{r}_{ij}-r_{ij} | 
\end{equation}
where $R^{Test}$ is the test set.
Unlike MAE, the RMSE emphasizes more significant errors by first squaring each error value, averaging the squared errors, then taking the square root of the average:
\begin{equation}
	RMSE = \sqrt{\frac{1}{|R^{Test}|} \sum_{r_{ij} \in R^{Test}} \left( \hat{r}_{ij}-r_{ij} \right)^2}.
\end{equation}
To aggregate the performance over all datasets, we first rank the methods according to their performance on each dataset with the convention that lower ranks are better then report the mean rank over all datasets.

\subsection{Competing methods}
The proposed method is experimentally compared against classical baselines as well as state-of-the-art algorithms to assess its performance. The following methods have been used in this experiment:
\begin{itemize}
	\item \textbf{ItemKNN:} An item-based CF method that computes the average rating of the $k$ most similar neighbors to the target item. The similarity between the items is based on Pearson correlation Coefficient. Since it serves as a baseline, this method was used un-tuned with a constant number of neighbors, $k = 25$.
	
	\item \textbf{SVD++:} An efficient algorithm that extends the classical SVD model by including implicit feedback and mixing the strengths of latent factor models and neighborhood models \cite{Koren2008SVDPP}. 
	
	\item \textbf{PMF:} An efficient matrix factorization model that factorizes the explicit user-item rating matrix as a product of two lower-rank users and items matrices to be used for predicting ratings \cite{salakhutdinov2008PMF}. 
	\item \textbf{BiasedMF:} An efficient biased matrix factorization model that extends factorizing explicit user-item rating matrices by introducing users and item biases \cite{koren2009BiasedMF}.
\end{itemize}

\subsection{Implementation}
All the variants of the proposed method are implemented using C++. Since an efficient and effective solver is essential for solving the optimization problems resulting from our models,  we use CG\_DESCENT \cite{Hager2006Algorithm}, which is a stable and well-tested code that implements a nonlinear conjugate gradient with a guaranteed descent method for unconstrained optimization \cite{Hager2005Anew}. To iteratively proceed towards the solution, the CG\_DESCENT solver requires, at each iteration, the value of the objective function and its gradient. For each of the proposed models, we implemented and tested two variations of the objective function, the $L_1$ and $L_2$ versions, as described in Section \ref{sec:proposed}. For the competing methods, a Java code on top of LibRec Java recommender system library \cite{guo2015librec} was used. 

\subsection{Parameter settings and tuning}
The experiment is conducted on 21 datasets. For each dataset, an extensive number of experiments were conducted to tune the hyper-parameters using grid search on a validation set. The dimensions $D$ for all proposed similarity-popularity models and factors $f$ in PMF, BiasedMF and SVD++ are selected from the set $\{5, 10, 20\}$. For the proposed similarity-popularity models, PMF, and BiasedMF, we tried the following combination of regularization $\lambda \in \{0.1, 0.01\}$. For PMF and BiasedMF we tried learning rate $\gamma \in \{0.1, 0.01\}$. For SPHM1, we included $\alpha \in \{2, 3, 4, ..., 8, 9\}$ in the grid search. For simplicity, we fixed $k$ in ItemKNN method to $k = 25$. All other parameters are kept to LibRec default values.

\subsection{Experimental results}
In the first experiment, we compare the performance of each variant of the proposed method. Namely, we compare the three proposed models, SPHM1, SPHM2, and SPDP, with the $L_1$ and $L_2$ objectives giving a total of six variants. Table \ref{tab:variants} reports the performance results in terms of RMSE and MAE of all six models on all datasets. The last row represents the mean rank of each model. The results show that SPHM2 with the $L_1$ objective function outperforms the other variants in terms of MAE in 17 of the 21 datasets, with a mean rank of 1.57. It is followed by SPHM1 with the $L_1$ objective, which comes in second with a mean rank of 2.52. 
Regarding RMSE, SPHM2 with the $L_2$ objective function and SPDP with the $L_1$ objective function perform better in 9 of the 21 datasets. However, SPHM2-L2 ranks better with an average of 1.62 compared to SPDP-L1 with a mean rank of 2.90. 

\begin{table*}[!t]
	\caption{Comparison of the different variants of the proposed method in terms of RMSE and MAE. The last row indicates the average rank over all datasets; the lower, the better.}
	\label{tab:variants}
	\centering
	\begin{tabular}{lrrrrrrrrrrrr}
		\toprule
		&           \multicolumn{6}{c}{RMSE}                                                               &                                                                    \multicolumn{6}{c}{MAE}                                                                      \\ \cmidrule(lr){2-7} \cmidrule(lr){8-13}
		& \rotatebox{90}{SPHM1-L1} & \rotatebox{90}{SPHM2-L1} & \rotatebox{90}{SPDP-L1} & \rotatebox{90}{SPHM1-L2} & \rotatebox{90}{SPHM2-L2} & \rotatebox{90}{SPDP-L2} & \rotatebox{90}{SPHM1-L1} & \rotatebox{90}{SPHM2-L1} & \rotatebox{90}{SPDP-L1} & \rotatebox{90}{SPHM1-L2} & \rotatebox{90}{SPHM2-L2} & \rotatebox{90}{SPDP-L2} \\  \cmidrule(lr){2-7} \cmidrule(lr){8-13}
		AmazonAA      &                    1.273 &                    1.291 &                   1.255 &                    1.253 &           \textbf{1.244} &                   1.401 &           \textbf{0.912} &                    0.914 &                   0.948 &                    0.988 &                    0.972 &                   1.063 \\
		AmazonAuto    &                    1.100 &                    1.102 &          \textbf{1.089} &                    1.109 &                    1.096 &                   1.104 &                    0.718 &           \textbf{0.703} &                   0.761 &                    0.772 &                    0.736 &                   0.766 \\
		AmazonBaby    &                    1.146 &                    1.154 &          \textbf{1.112} &                    1.158 &                    1.140 &                   1.183 &                    0.829 &           \textbf{0.800} &                   0.859 &                    0.898 &                    0.858 &                   0.891 \\
		AmazonDM      &                    0.794 &                    0.809 &                   0.792 &                    0.790 &           \textbf{0.784} &                   0.818 &                    0.467 &           \textbf{0.463} &                   0.515 &                    0.502 &                    0.484 &                   0.523 \\
		AmazonGGF     &                    1.089 &                    1.098 &          \textbf{1.080} &                    1.103 &                    1.087 &                   1.143 &                    0.730 &           \textbf{0.710} &                   0.789 &                    0.809 &                    0.769 &                   0.824 \\
		AmazonHPC     &                    1.139 &                    1.146 &          \textbf{1.119} &                    1.152 &                    1.136 &                   1.179 &                    0.782 &           \textbf{0.754} &                   0.830 &                    0.859 &                    0.817 &                   0.859 \\
		AmazonIV      &           \textbf{1.070} &                    1.081 &                   1.071 &                    1.081 &                    1.077 &                   1.085 &                    0.750 &           \textbf{0.736} &                   0.793 &                    0.803 &                    0.780 &                   0.805 \\
		AmazonMI      &                    1.009 &                    1.027 &          \textbf{0.993} &                    1.016 &                    1.005 &                   1.010 &                    0.662 &           \textbf{0.651} &                   0.709 &                    0.717 &                    0.682 &                   0.714 \\
		AmazonPLG     &                    1.182 &                    1.182 &          \textbf{1.165} &                    1.190 &                    1.180 &                   1.186 &                    0.819 &           \textbf{0.805} &                   0.859 &                    0.870 &                    0.830 &                   0.867 \\
		AmazonTHI     &                    1.097 &                    1.096 &          \textbf{1.076} &                    1.109 &                    1.094 &                   1.107 &                    0.747 &           \textbf{0.722} &                   0.783 &                    0.805 &                    0.764 &                   0.796 \\
		AmazonVG      &                    1.132 &                    1.135 &          \textbf{1.105} &                    1.136 &                    1.129 &                   1.186 &                    0.818 &           \textbf{0.801} &                   0.859 &                    0.876 &                    0.851 &                   0.899 \\
		Anime         &                    1.175 &                    1.168 &                   1.172 &                    1.144 &           \textbf{1.138} &                   1.142 &                    0.871 &                    0.866 &                   0.860 &                    0.867 &                    0.863 &          \textbf{0.857} \\
		Book-Crossing &                    3.790 &                    3.745 &                   3.761 &                    3.450 &           \textbf{3.416} &                   3.819 &                    2.403 &           \textbf{2.392} &                   2.568 &                    2.625 &                    2.767 &                   2.797 \\
		CiaoDVD       &                    0.986 &                    0.996 &          \textbf{0.957} &                    0.984 &                    0.982 &                   1.011 &                    0.734 &           \textbf{0.730} &                   0.746 &                    0.761 &                    0.752 &                   0.779 \\
		Epinions      &                    1.087 &                    1.078 &                   1.165 &                    1.069 &           \textbf{1.060} &                   1.200 &                    0.795 &           \textbf{0.782} &                   0.859 &                    0.836 &                    0.819 &                   0.901 \\
		FilmTrust     &                    0.805 &                    0.805 &                   0.868 &                    0.796 &           \textbf{0.791} &                   0.862 &                    0.607 &           \textbf{0.603} &                   0.644 &                    0.618 &                    0.613 &                   0.646 \\
		Food.com      &                    0.989 &                    0.988 &                   0.967 &                    0.960 &           \textbf{0.938} &                   1.062 &           \textbf{0.438} &           \textbf{0.438} &                   0.505 &                    0.585 &                    0.558 &                   0.621 \\
		ML100K        &                    0.932 &                    0.930 &                   0.989 &                    0.915 &           \textbf{0.912} &                   0.948 &                    0.728 &           \textbf{0.724} &                   0.755 &                    0.726 &           \textbf{0.724} &                   0.734 \\
		ML1M          &                    0.872 &                    0.872 &                   0.893 &           \textbf{0.852} &                    0.853 &                   0.867 &                    0.679 &                    0.678 &                   0.679 &           \textbf{0.674} &                    0.676 &                   0.678 \\
		YahooMovies   &                    3.062 &                    3.064 &                   3.221 &                    2.944 &           \textbf{2.941} &                   3.279 &                    2.070 &           \textbf{2.058} &                   2.185 &                    2.172 &                    2.200 &                   2.295 \\
		YahooMusic    &                    1.670 &                    1.666 &                   1.370 &           \textbf{1.178} &                    1.190 &                   1.316 &                    1.186 &                    1.190 &                   0.970 &           \textbf{0.955} &                    0.989 &                   0.958 \\ \cmidrule(lr){2-7} \cmidrule(lr){8-13}
		Mean Rank     &                     3.76 &                     4.14 &                    2.90 &                     3.38 &            \textbf{1.62} &                    5.19 &                     2.52 &            \textbf{1.57} &                    4.00 &                     4.52 &                     3.29 &                    5.10 \\ \bottomrule
		&                          &                          &
	\end{tabular}
\end{table*}

\begin{figure*}[!t]
	\centering
	\includegraphics[width=0.4\linewidth]{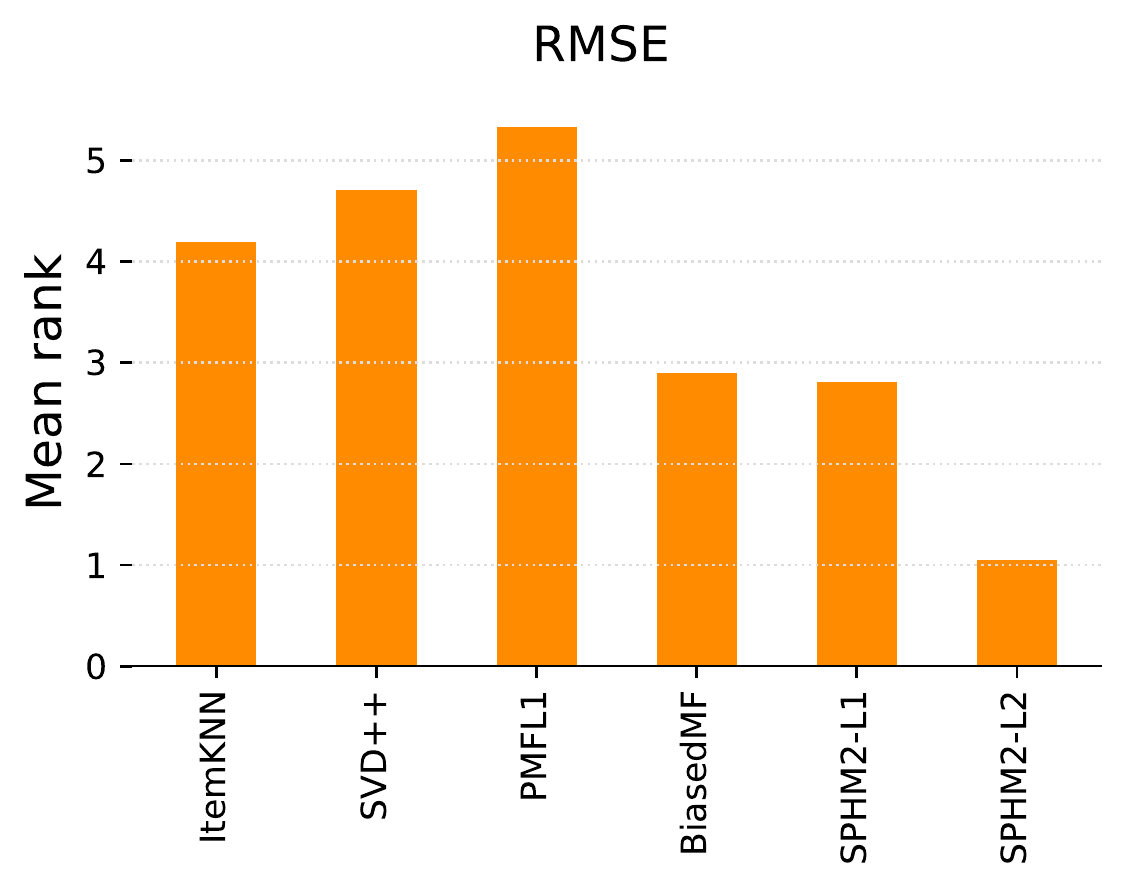}
	\quad
	\includegraphics[width=0.4\linewidth]{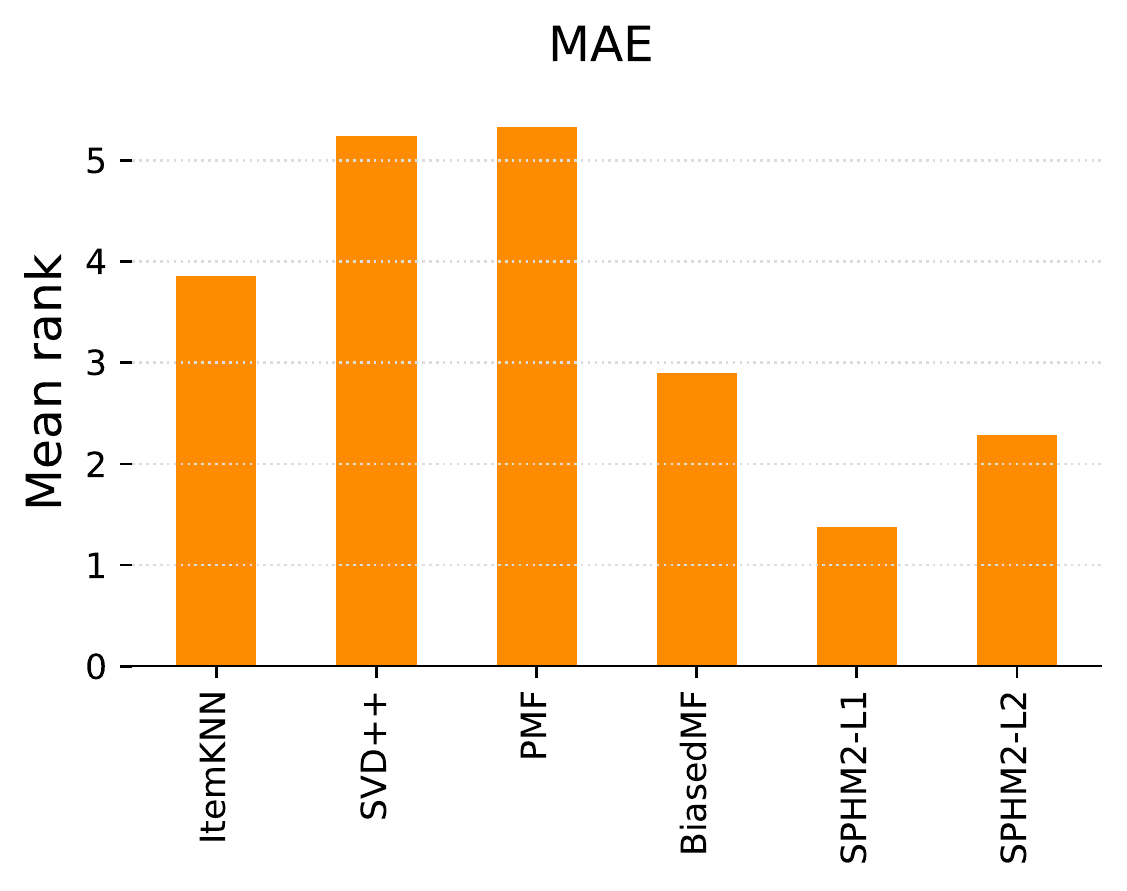}
	\caption{Mean rank comparison over all datasets in terms of RMSE and MAE.}
	\label{fig:competing}
\end{figure*}

The top-performing methods in terms of MAE and RMSE, respectively SPHM2-L1 and SPHM2-L2, are picked for the second experiment, where we compare their performance against state-of-the-art recommendation methods. Table \ref{tab:competing} reports the RMSE and MAE results of SPHM2-L1 and SPHM2-L2 and the competing methods. The mean ranks over all datasets are reported in the last row and plotted for convenience in Figure \ref{fig:competing}.

The results show that SPHM2-L1 outperforms the other methods in 18 of 21 datasets in terms of MAE with a mean rank of 1.38, followed by SPHM2-L2 with a mean rank of 2.29. SPHM2-L2 outperforms other methods in 20 of 21 datasets in terms of RMSE with a mean rank of 1.05, followed by SPHM2-L1 with a mean rank of 2.81.

\begin{table*}[!t]
	\caption{Comparison of the proposed approach against competing methods in terms of RMSE and MAE. The last row indicates the average rank over all datasets; the lower, the better.}
	\label{tab:competing}
	\centering
	\begin{tabular}{lrrrrrrrrrrrr}
		\toprule
		&                                                                 \multicolumn{6}{c}{RMSE}                                                                 &                                                                \multicolumn{6}{c}{MAE}                                                                 \\
		\cmidrule(lr){2-7} \cmidrule(lr){8-13}            & \rotatebox{90}{ItemKNN} & \rotatebox{90}{SVD++} & \rotatebox{90}{PMF} & \rotatebox{90}{BiasedMF} & \rotatebox{90}{SPHM2-L1} & \rotatebox{90}{SPHM2-L2} & \rotatebox{90}{ItemKNN} & \rotatebox{90}{SVD++} & \rotatebox{90}{PMF} & \rotatebox{90}{BiasedMF} & \rotatebox{90}{SPHM2-L1} & \rotatebox{90}{SPHM2-L2} \\
		\cmidrule(lr){2-7} \cmidrule(lr){8-13}
		AmazonAA  &                   1.379 &                 1.378 &                 1.378 &                    1.323 &                    1.291 &           \textbf{1.244} &                   1.026 &                 1.121 &               1.122 &                    0.978 &           \textbf{0.914} &                    0.972 \\
		AmazonAuto                                        &                   1.109 &                 1.108 &                 1.160 &                    1.101 &                    1.102 &           \textbf{1.096} &                   0.849 &                 0.852 &               0.878 &                    0.757 &           \textbf{0.703} &                    0.736 \\
		AmazonBaby                                        &                   1.187 &                 1.171 &                 1.501 &                    1.169 &                    1.154 &           \textbf{1.140} &                   0.931 &                 0.936 &               1.082 &                    0.858 &           \textbf{0.800} &                    0.858 \\
		AmazonDM                                          &                   0.876 &                 0.909 &                 1.280 &                    0.794 &                    0.809 &           \textbf{0.784} &                   0.660 &                 0.688 &               0.873 &                    0.508 &           \textbf{0.463} &                    0.484 \\
		AmazonGGF                                         &                   1.131 &                 1.139 &                 1.828 &                    1.119 &                    1.098 &           \textbf{1.087} &                   0.888 &                 0.904 &               1.283 &                    0.792 &           \textbf{0.710} &                    0.769 \\
		AmazonHPC                                         &                   1.183 &                 1.185 &                 1.622 &                    1.161 &                    1.146 &           \textbf{1.136} &                   0.923 &                 0.939 &               1.123 &                    0.825 &           \textbf{0.754} &                    0.817 \\
		AmazonIV                                          &                   1.198 &                 1.211 &                 1.457 &           \textbf{1.075} &                    1.081 &                    1.077 &                   0.903 &                 0.961 &               1.003 &                    0.766 &           \textbf{0.736} &                    0.780 \\
		AmazonMI                                          &                   1.023 &                 1.019 &                 1.057 &                    1.007 &                    1.027 &           \textbf{1.005} &                   0.794 &                 0.796 &               0.813 &                    0.702 &           \textbf{0.651} &                    0.682 \\
		AmazonPLG                                         &                   1.237 &                 1.237 &                 1.690 &                    1.184 &                    1.182 &           \textbf{1.180} &                   0.974 &                 0.979 &               1.263 &                    0.847 &           \textbf{0.805} &                    0.830 \\
		AmazonTHI                                         &                   1.126 &                 1.124 &                 1.576 &                    1.098 &                    1.096 &           \textbf{1.094} &                   0.874 &                 0.877 &               1.155 &                    0.771 &           \textbf{0.722} &                    0.764 \\
		AmazonVG                                          &                   1.224 &                 1.232 &                 1.336 &                    1.151 &                    1.135 &           \textbf{1.129} &                   0.926 &                 0.970 &               0.973 &                    0.842 &           \textbf{0.801} &                    0.851 \\
		Anime                                             &                   1.180 &                 1.572 &                 1.197 &                    1.146 &                    1.168 &           \textbf{1.138} &                   0.886 &                 1.232 &               0.900 &           \textbf{0.859} &                    0.866 &                    0.863 \\
		Book-Crossing                                     &                   3.767 &                 3.801 &                 3.804 &                    3.724 &                    3.745 &           \textbf{3.416} &                   3.184 &                 3.483 &               3.484 &                    2.782 &           \textbf{2.392} &                    2.767 \\
		CiaoDVD                                           &                   1.094 &                 1.103 &                 1.640 &                    0.984 &                    0.996 &           \textbf{0.982} &                   0.832 &                 0.844 &               1.189 &                    0.742 &           \textbf{0.730} &                    0.752 \\
		Epinions                                          &                   1.174 &                 1.204 &                 1.386 &                    1.142 &                    1.078 &           \textbf{1.060} &                   0.885 &                 0.916 &               1.030 &                    0.851 &           \textbf{0.782} &                    0.819 \\
		FilmTrust                                         &                   0.811 &                 0.918 &                 0.939 &                    0.841 &                    0.805 &           \textbf{0.791} &                   0.613 &                 0.715 &               0.681 &                    0.628 &           \textbf{0.603} &                    0.613 \\
		Food.com                                          &                   0.986 &                 0.956 &                 0.956 &                    1.067 &                    0.988 &           \textbf{0.938} &                   0.623 &                 0.633 &               0.633 &                    0.639 &           \textbf{0.438} &                    0.558 \\
		ML100K                                            &                   0.951 &                 1.126 &                 0.946 &                    0.950 &                    0.930 &           \textbf{0.912} &                   0.746 &                 0.945 &               0.733 &                    0.735 &           \textbf{0.724} &           \textbf{0.724} \\
		ML1M                                              &                   0.899 &                 1.117 &                 0.881 &                    0.866 &                    0.872 &           \textbf{0.853} &                   0.705 &                 0.934 &               0.688 &           \textbf{0.675} &                    0.678 &                    0.676 \\
		YahooMovies                                       &                   3.065 &                 3.603 &                 3.603 &                    3.346 &                    3.064 &           \textbf{2.941} &                   2.186 &                 2.848 &               2.848 &                    2.303 &           \textbf{2.058} &                    2.200 \\
		YahooMusic                                        &                   1.221 &                 1.565 &                 1.261 &                    1.281 &                    1.666 &           \textbf{1.190} &                   0.943 &                 1.409 &      \textbf{0.924} &                    0.937 &                    1.190 &                    0.989 \\
		\cmidrule(lr){2-7} \cmidrule(lr){8-13}
		Mean Rank &                    4.19 &                  4.71 &                  5.33 &                     2.90 &                     2.81 &            \textbf{1.05} &                    3.86 &                  5.24 &                5.33 &                     2.90 &            \textbf{1.38} &                     2.29 \\ \bottomrule
	\end{tabular}
\end{table*}

\begin{figure*}[!t]
	\centering
	\includegraphics[width=0.4\linewidth]{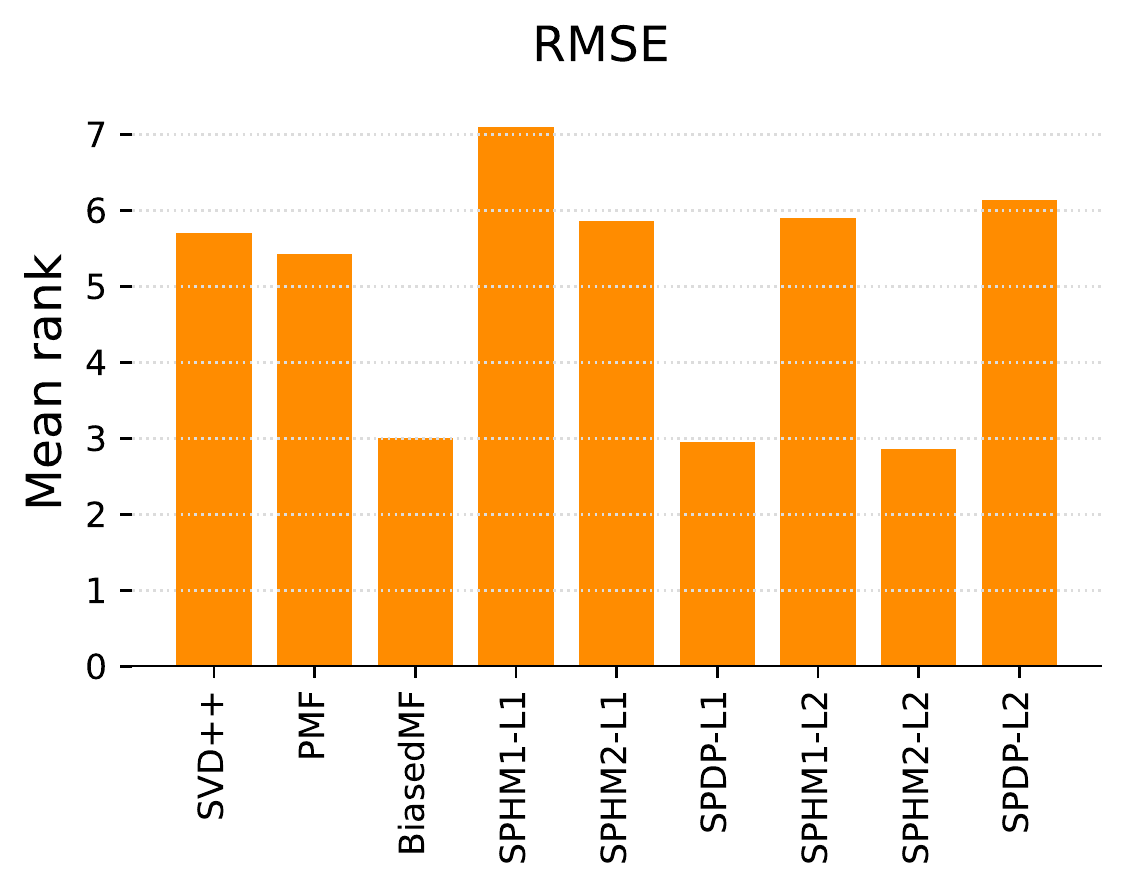}
	\quad
	\includegraphics[width=0.4\linewidth]{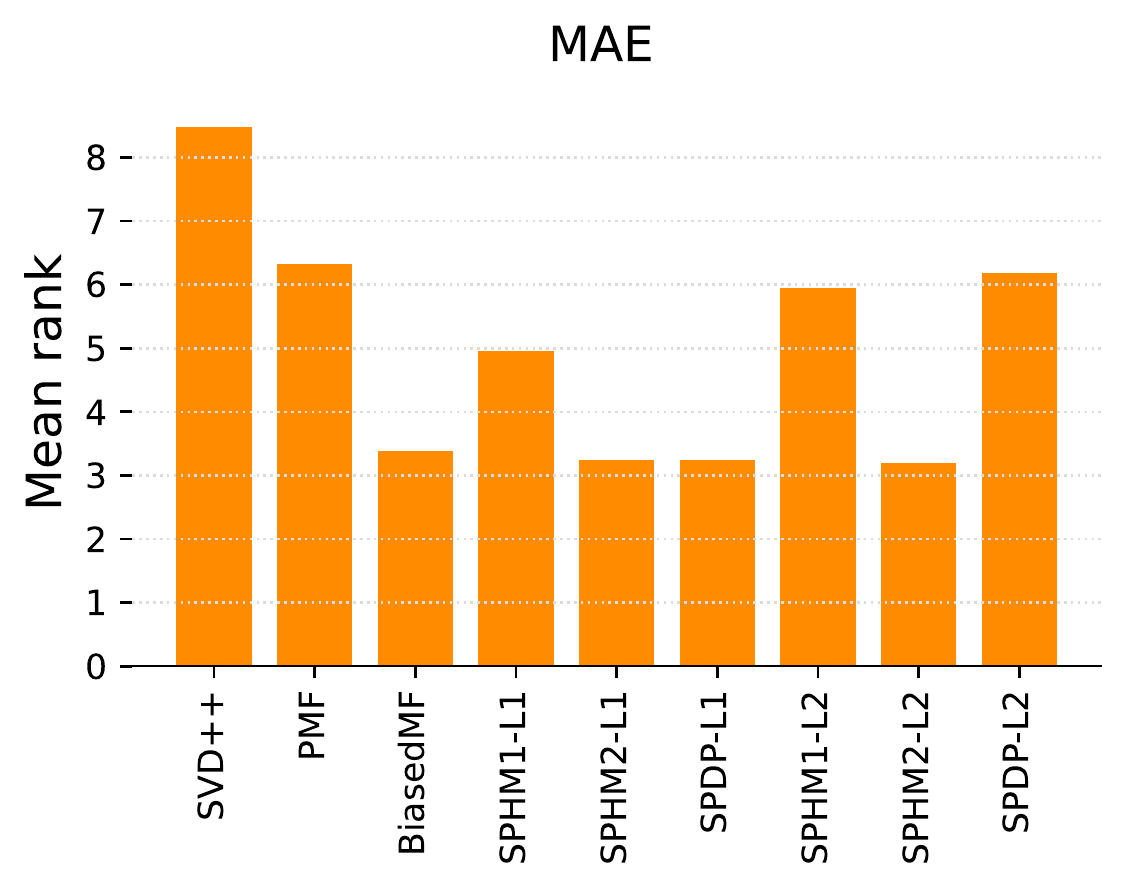}
	\caption{Mean rank comparison over all datasets in terms of RMSE and MAE in 3D.}
	\label{fig:competing-3D}
\end{figure*}

We conduct an additional experiment to evaluate the data visualization prospect of the proposed approach against the competing methods. Data visualization requires embedding users and items in a two or three-dimensional space, allowing data exploration and interpretation. The experiment consists in training all methods on three dimensions except obviously for ItemKNN, which does not assign coordinates to items and users. The resulting mean ranks based on RMSE and MAE depicted in Figure \ref{fig:competing-3D} show that the proposed method SPHM2-L2 performs better in terms of MAE and RMSE, followed closely by SPDP-L1 then BiasedMF.

\section{Conclusion and Future Work} \label{sec:conclusion}
This work introduced a novel recommendation method that utilizes a similarity-popularity complex network model to generate recommendations. The network, which has users and items as nodes, is first constructed from observed ratings and then used to predict unseen ratings. We explored the effectiveness of dot-product and Euclidean distance-based similarity with $L_1$ and $L_2$ norms for error cost and regularization. We implemented the proposed models, conducted extensive experiments on multiple datasets from diverse domains, and demonstrated the superiority of the proposed approach against state-of-the-art recommendation methods. We also showed that the proposed approach outperforms existing methods in low dimensions, proving the method's effectiveness for data visualization and exploration. 

The proposed similarity-popularity model is a starting point in the direction of applying complex network latent spaces as models that guide the recommendations process. Additional similarity-popularity models other than the hidden metric space model and dot-product can be examined to assess their suitability for recommendation systems. Furthermore, other objective functions can be tested, especially one that combines both L1 and L2, such as an elastic net. 

Another direction is to apply the proposed method to other forms of the recommendation problem, such as item ranking, context-aware, and sequence-aware recommendation tasks. We also propose to investigate adding content-based features along with similarity and popularity to tackle further recommendation systems challenges, such as recommendation diversification and the cold start problem.

\ifCLASSOPTIONcompsoc
  \section*{Acknowledgments}
        This research work is supported by the Research Center, CCIS, King Saud University, Riyadh, Saudi Arabia.
\else
  \section*{Acknowledgment}
\fi

\ifCLASSOPTIONcaptionsoff
  \newpage
\fi

\IEEEtriggeratref{48}
\bibliographystyle{IEEEtran}

\begin{thebibliography}{10}
	\providecommand{\url}[1]{#1}
	\csname url@samestyle\endcsname
	\providecommand{\newblock}{\relax}
	\providecommand{\bibinfo}[2]{#2}
	\providecommand{\BIBentrySTDinterwordspacing}{\spaceskip=0pt\relax}
	\providecommand{\BIBentryALTinterwordstretchfactor}{4}
	\providecommand{\BIBentryALTinterwordspacing}{\spaceskip=\fontdimen2\font plus
		\BIBentryALTinterwordstretchfactor\fontdimen3\font minus
		\fontdimen4\font\relax}
	\providecommand{\BIBforeignlanguage}[2]{{%
			\expandafter\ifx\csname l@#1\endcsname\relax
			\typeout{** WARNING: IEEEtran.bst: No hyphenation pattern has been}%
			\typeout{** loaded for the language `#1'. Using the pattern for}%
			\typeout{** the default language instead.}%
			\else
			\language=\csname l@#1\endcsname
			\fi
			#2}}
	\providecommand{\BIBdecl}{\relax}
	\BIBdecl
	
	\bibitem{Bobadilla2013Survey}
	\BIBentryALTinterwordspacing
	J.~Bobadilla, F.~Ortega, A.~Hernando, and A.~Guti{\'{e}}rrez, ``Recommender
	systems survey,'' \emph{Knowledge-Based Systems}, vol.~46, pp. 109--132,
	2013. [Online]. Available:
	\url{http://www.sciencedirect.com/science/article/pii/S0950705113001044}
	\BIBentrySTDinterwordspacing
	
	\bibitem{Lu2015Survey}
	\BIBentryALTinterwordspacing
	J.~Lu, D.~Wu, M.~Mao, W.~Wang, and G.~Zhang, ``Recommender system application
	developments: A survey,'' \emph{Decision Support Systems}, vol.~74, pp.
	12--32, 2015. [Online]. Available:
	\url{http://www.sciencedirect.com/science/article/pii/S0167923615000627}
	\BIBentrySTDinterwordspacing
	
	\bibitem{handbook2015chapter2}
	\BIBentryALTinterwordspacing
	X.~Ning, C.~Desrosiers, and G.~Karypis, ``A comprehensive survey of
	neighborhood-based recommendation methods,'' in \emph{Recommender Systems
		Handbook}, F.~Ricci, L.~Rokach, and B.~Shapira, Eds.\hskip 1em plus 0.5em
	minus 0.4em\relax Boston, MA: Springer US, 2015, pp. 37--76. [Online].
	Available: \url{https://doi.org/10.1007/978-1-4899-7637-6_2}
	\BIBentrySTDinterwordspacing
	
	\bibitem{goyal2018graph}
	\BIBentryALTinterwordspacing
	P.~Goyal and E.~Ferrara, ``Graph embedding techniques, applications, and
	performance: A survey,'' \emph{Knowledge-Based Systems}, vol. 151, pp.
	78--94, 2018. [Online]. Available:
	\url{http://www.sciencedirect.com/science/article/pii/S0950705118301540}
	\BIBentrySTDinterwordspacing
	
	\bibitem{cai2018comprehensive}
	\BIBentryALTinterwordspacing
	H.~Cai, V.~W. Zheng, and K.~C. Chang, ``A comprehensive survey of graph
	embedding: Problems, techniques, and applications,'' \emph{IEEE Transactions
		on Knowledge \& Data Engineering}, vol.~30, no.~9, pp. 1616--1637, Sept.
	2018. [Online]. Available:
	\url{doi.ieeecomputersociety.org/10.1109/TKDE.2018.2807452}
	\BIBentrySTDinterwordspacing
	
	\bibitem{huang2007analyzing}
	\BIBentryALTinterwordspacing
	Z.~Huang, D.~D. Zeng, and H.~Chen, ``Analyzing consumer-product graphs:
	Empirical findings and applications in recommender systems,''
	\emph{Management science}, vol.~53, no.~7, pp. 1146--1164, 2007. [Online].
	Available: \url{https://doi.org/10.1287/mnsc.1060.0619}
	\BIBentrySTDinterwordspacing
	
	\bibitem{zhou2007bipartite}
	\BIBentryALTinterwordspacing
	T.~Zhou, J.~Ren, M.~c.~v. Medo, and Y.-C. Zhang, ``Bipartite network projection
	and personal recommendation,'' \emph{Phys. Rev. E}, vol.~76, p. 046115, Oct
	2007. [Online]. Available:
	\url{https://link.aps.org/doi/10.1103/PhysRevE.76.046115}
	\BIBentrySTDinterwordspacing
	
	\bibitem{zanin2008complex}
	\BIBentryALTinterwordspacing
	M.~Zanin, P.~Cano, J.~M. Buld\'{u}, and O.~Celma, ``Complex networks in
	recommendation systems,'' in \emph{Proceedings of the 2Nd WSEAS International
		Conference on Computer Engineering and Applications}, ser. CEA'08.\hskip 1em
	plus 0.5em minus 0.4em\relax Stevens Point, Wisconsin, USA: World Scientific
	and Engineering Academy and Society (WSEAS), 2008, pp. 120--124. [Online].
	Available: \url{http://dl.acm.org/citation.cfm?id=1373936.1373955}
	\BIBentrySTDinterwordspacing
	
	\bibitem{boguna2010sustaining}
	M.~Bogun{\'a}, F.~Papadopoulos, and D.~Krioukov, ``Sustaining the internet with
	hyperbolic mapping,'' \emph{Nature communications}, vol.~1, p.~62, 2010.
	
	\bibitem{costa2011complexsurvey}
	\BIBentryALTinterwordspacing
	L.~d.~F. Costa, O.~N. Oliveira~Jr, G.~Travieso, F.~A. Rodrigues, P.~R.
	Villas~Boas, L.~Antiqueira, M.~P. Viana, and L.~E. Correa~Rocha, ``Analyzing
	and modeling real-world phenomena with complex networks: a survey of
	applications,'' \emph{Advances in Physics}, vol.~60, no.~3, pp. 329--412,
	2011. [Online]. Available: \url{https://doi.org/10.1080/00018732.2011.572452}
	\BIBentrySTDinterwordspacing
	
	\bibitem{newman2001collaboration}
	\BIBentryALTinterwordspacing
	M.~E. Newman, ``The structure of scientific collaboration networks,''
	\emph{Proceedings of the National Academy of Sciences}, vol.~98, no.~2, pp.
	404--409, 2001. [Online]. Available:
	\url{http://www.pnas.org/content/98/2/404}
	\BIBentrySTDinterwordspacing
	
	\bibitem{barabasi2000scale}
	A.-L. Barab{\'a}si, R.~Albert, and H.~Jeong, ``Scale-free characteristics of
	random networks: the topology of the world-wide web,'' \emph{Physica A:
		Statistical Mechanics and its Applications}, vol. 281, no. 1-4, pp. 69--77,
	2000.
	
	\bibitem{williams2002food}
	\BIBentryALTinterwordspacing
	R.~J. Williams, E.~L. Berlow, J.~A. Dunne, A.-L. Barab{\'a}si, and N.~D.
	Martinez, ``Two degrees of separation in complex food webs,''
	\emph{Proceedings of the National Academy of Sciences}, vol.~99, no.~20, pp.
	12\,913--12\,916, 2002. [Online]. Available:
	\url{http://www.pnas.org/content/99/20/12913}
	\BIBentrySTDinterwordspacing
	
	\bibitem{yuan2010small-worldness}
	\BIBentryALTinterwordspacing
	W.~Yuan, D.~Guan, Y.-K. Lee, S.~Lee, and S.~J. Hur, ``Improved trust-aware
	recommender system using small-worldness of trust networks,''
	\emph{Knowledge-Based Systems}, 2010. [Online]. Available:
	\url{http://www.sciencedirect.com/science/article/pii/S095070511000002X}
	\BIBentrySTDinterwordspacing
	
	\bibitem{trifunovic2010social}
	S.~Trifunovic, F.~Legendre, and C.~Anastasiades, ``Social trust in
	opportunistic networks,'' in \emph{2010 INFOCOM IEEE Conference on Computer
		Communications Workshops}.\hskip 1em plus 0.5em minus 0.4em\relax IEEE, 2010,
	pp. 1--6.
	
	\bibitem{zhou2011emergence}
	\BIBentryALTinterwordspacing
	T.~Zhou, M.~Medo, G.~Cimini, Z.-K. Zhang, and Y.-C. Zhang, ``Emergence of
	scale-free leadership structure in social recommender systems,'' \emph{PLOS
		ONE}, vol.~6, pp. 1--6, 07 2011. [Online]. Available:
	\url{https://doi.org/10.1371/journal.pone.0020648}
	\BIBentrySTDinterwordspacing
	
	\bibitem{cano2006music}
	\BIBentryALTinterwordspacing
	P.~Cano, O.~Celma, M.~Koppenberger, and J.~M. Buldu, ``Topology of music
	recommendation networks,'' \emph{Chaos: An Interdisciplinary Journal of
		Nonlinear Science}, vol.~16, no.~1, p. 013107, 2006. [Online]. Available:
	\url{https://doi.org/10.1063/1.2137622}
	\BIBentrySTDinterwordspacing
	
	\bibitem{yuan2013recommender}
	\BIBentryALTinterwordspacing
	W.~Yuan, D.~Guan, L.~Shu, and J.~Niu, ``Recommender searching mechanism for
	trust-aware recommender systems in internet of things,'' \emph{automatika},
	vol.~54, no.~4, pp. 427--437, 2013. [Online]. Available:
	\url{https://doi.org/10.7305/automatika.54-4.416}
	\BIBentrySTDinterwordspacing
	
	\bibitem{ERmodel1959}
	\BIBentryALTinterwordspacing
	P.~Erd{\H o}s and A.~R{\'e}nyi, ``On random graphs, i,'' \emph{Publicationes
		Mathematicae (Debrecen)}, vol.~6, pp. 290--297, 1959. [Online]. Available:
	\url{http://www.renyi.hu/~p_erdos/Erdos.html#1959-11}
	\BIBentrySTDinterwordspacing
	
	\bibitem{barabasi2009decade}
	\BIBentryALTinterwordspacing
	A.-L. Barab{\'a}si, ``Scale-free networks: a decade and beyond,''
	\emph{Science}, vol. 325, no. 5939, pp. 412--413, 2009. [Online]. Available:
	\url{http://science.sciencemag.org/content/325/5939/412}
	\BIBentrySTDinterwordspacing
	
	\bibitem{wang2003complex}
	X.~F. Wang and G.~Chen, ``Complex networks: small-world, scale-free and
	beyond,'' \emph{IEEE Circuits and Systems Magazine}, vol.~3, no.~1, pp.
	6--20, 2003.
	
	\bibitem{WSmodel998}
	D.~J. Watts and S.~H. Strogatz, ``Collective dynamics of
	‘small-world’networks,'' \emph{Nature}, vol. 393, no. 6684, p. 440, 1998.
	
	\bibitem{BAmodel1999}
	\BIBentryALTinterwordspacing
	A.-L. Barab{\'a}si and R.~Albert, ``Emergence of scaling in random networks,''
	\emph{Science}, vol. 286, no. 5439, pp. 509--512, 1999. [Online]. Available:
	\url{http://science.sciencemag.org/content/286/5439/509}
	\BIBentrySTDinterwordspacing
	
	\bibitem{serrano2008HMS}
	\BIBentryALTinterwordspacing
	M.~A. Serrano, D.~Krioukov, and M.~Bogun{\'a}, ``Self-similarity of complex
	networks and hidden metric spaces,'' \emph{Physical review letters}, vol.
	100, no.~7, p. 078701, 2008. [Online]. Available:
	\url{https://link.aps.org/doi/10.1103/PhysRevLett.100.078701}
	\BIBentrySTDinterwordspacing
	
	\bibitem{boguna2009navigabilityHMS}
	\BIBentryALTinterwordspacing
	M.~Bogun{\'a}, D.~Krioukov, and K.~C. Claffy, ``Navigability of complex
	networks,'' \emph{Nature Physics}, vol.~5, no.~1, p.~74, 2009. [Online].
	Available:
	\url{https://www.nature.com/articles/nphys1130#supplementary-information}
	\BIBentrySTDinterwordspacing
	
	\bibitem{muscoloni2018nonuniform}
	\BIBentryALTinterwordspacing
	A.~Muscoloni and C.~V. Cannistraci, ``A nonuniform popularity-similarity
	optimization (npso) model to efficiently generate realistic complex networks
	with communities,'' \emph{New Journal of Physics}, vol.~20, no.~5, p. 052002,
	2018. [Online]. Available: \url{https://doi.org/10.1088/1367-2630/aac06f}
	\BIBentrySTDinterwordspacing
	
	\bibitem{boguna2021network}
	\BIBentryALTinterwordspacing
	M.~Boguna, I.~Bonamassa, M.~De~Domenico, S.~Havlin, D.~Krioukov, and
	M.~Serrano, ``Network geometry,'' \emph{Nature Reviews Physics}, vol.~3,
	no.~2, pp. 114--135, 2021. [Online]. Available:
	\url{https://doi.org/10.1038/s42254-020-00264-4}
	\BIBentrySTDinterwordspacing
	
	\bibitem{Papadopoulos2015NetworkMappingbyReplayingHyperbolicGrowth}
	F.~Papadopoulos, C.~Psomas, and D.~Krioukov, ``Network mapping by replaying
	hyperbolic growth,'' \emph{IEEE/ACM Transactions on Networking (TON)},
	vol.~23, no.~1, pp. 198--211, 2015.
	
	\bibitem{kerrache16}
	R.~Alharbi, B.~Hafida, and S.~Kerrache, ``Scalable link prediction in complex
	networks using a type of geodesic distance,'' in \emph{Asia Multi Conference
		on Mathematical Modelling and Computer Simulatio}, 2016.
	
	\bibitem{kerrache20}
	S.~Kerrache, R.~Alharbi, and H.~Benhidour, ``\BIBforeignlanguage{eng}{A
		scalable similarity-popularity link prediction method},''
	\emph{\BIBforeignlanguage{eng}{Scientific reports}}, vol.~10, no.~1, pp.
	6394--6394, 2020.
	
	\bibitem{aggarwal1999horting}
	C.~C. Aggarwal, J.~L. Wolf, K.-L. Wu, and P.~S. Yu, ``Horting hatches an egg: A
	new graph-theoretic approach to collaborative filtering,'' in
	\emph{Proceedings of the fifth ACM SIGKDD international conference on
		Knowledge discovery and data mining}.\hskip 1em plus 0.5em minus 0.4em\relax
	ACM, 1999, pp. 201--212.
	
	\bibitem{Shi2014Survey}
	\BIBentryALTinterwordspacing
	Y.~Shi, M.~Larson, and A.~Hanjalic, ``Collaborative filtering beyond the
	user-item matrix : A survey of the state of the art and future challenges,''
	\emph{ACM Computing Surveys (CSUR)}, vol.~47, no.~1, pp. 1--45, 2014.
	[Online]. Available: \url{http://doi.acm.org/10.1145/2556270}
	\BIBentrySTDinterwordspacing
	
	\bibitem{lee2015escaping}
	\BIBentryALTinterwordspacing
	K.~Lee and K.~Lee, ``Escaping your comfort zone: A graph-based recommender
	system for finding novel recommendations among relevant items,'' \emph{Expert
		Systems with Applications}, vol.~42, no.~10, pp. 4851--4858, 2015. [Online].
	Available:
	\url{http://www.sciencedirect.com/science/article/pii/S0957417414004308}
	\BIBentrySTDinterwordspacing
	
	\bibitem{khoshneshin2010collaborative}
	\BIBentryALTinterwordspacing
	M.~Khoshneshin and W.~N. Street, ``Collaborative filtering via euclidean
	embedding,'' in \emph{Proceedings of the Fourth ACM Conference on Recommender
		Systems}, ser. RecSys '10.\hskip 1em plus 0.5em minus 0.4em\relax New York,
	NY, USA: ACM, 2010, pp. 87--94. [Online]. Available:
	\url{http://doi.acm.org/10.1145/1864708.1864728}
	\BIBentrySTDinterwordspacing
	
	\bibitem{walter2008trust}
	\BIBentryALTinterwordspacing
	F.~E. Walter, S.~Battiston, and F.~Schweitzer, ``A model of a trust-based
	recommendation system on a social network,'' \emph{Autonomous Agents and
		Multi-Agent Systems}, vol.~16, no.~1, pp. 57--74, 2008. [Online]. Available:
	\url{https://doi.org/10.1007/s10458-007-9021-x}
	\BIBentrySTDinterwordspacing
	
	\bibitem{azadjalal2017trust}
	\BIBentryALTinterwordspacing
	M.~M. Azadjalal, P.~Moradi, A.~Abdollahpouri, and M.~Jalili, ``A trust-aware
	recommendation method based on pareto dominance and confidence concepts,''
	\emph{Knowledge-Based Systems}, vol. 116, pp. 130--143, 2017. [Online].
	Available:
	\url{http://www.sciencedirect.com/science/article/pii/S0950705116304208}
	\BIBentrySTDinterwordspacing
	
	\bibitem{Lu2011LinkPredictioninComplexNetworks}
	L.~L{\"u} and T.~Zhou, ``Link prediction in complex networks: A survey,''
	\emph{Physica A: Statistical Mechanics and its Applications}, vol. 390,
	no.~6, pp. 1150--1170, 2011.
	
	\bibitem{wan2016modeling}
	M.~Wan and J.~McAuley, ``Modeling ambiguity, subjectivity, and diverging
	viewpoints in opinion question answering systems,'' in \emph{2016 IEEE 16th
		international conference on data mining (ICDM)}.\hskip 1em plus 0.5em minus
	0.4em\relax IEEE, 2016, pp. 489--498.
	
	\bibitem{Anime}
	\BIBentryALTinterwordspacing
	(2016) Anime recommendations database. [Online]. Available:
	\url{https://www.kaggle.com/datasets/CooperUnion/anime-recommendations-database}
	\BIBentrySTDinterwordspacing
	
	\bibitem{ziegler2005improving}
	C.-N. Ziegler, S.~M. McNee, J.~A. Konstan, and G.~Lausen, ``Improving
	recommendation lists through topic diversification,'' in \emph{Proceedings of
		the 14th international conference on World Wide Web}, 2005, pp. 22--32.
	
	\bibitem{guo2014etaf}
	G.~Guo, J.~Zhang, D.~Thalmann, and N.~Yorke-Smith, ``Etaf: An extended trust
	antecedents framework for trust prediction,'' in \emph{Proceedings of the
		2014 International Conference on Advances in Social Networks Analysis and
		Mining (ASONAM)}, 2014, pp. 540--547.
	
	\bibitem{massa2008trustlet}
	P.~Massa, K.~Souren, M.~Salvetti, and D.~Tomasoni, ``Trustlet, open research on
	trust metrics,'' \emph{Scalable Computing: Practice and Experience}, vol.~9,
	no.~4, 2008.
	
	\bibitem{guo2013novel}
	G.~Guo, J.~Zhang, and N.~Yorke-Smith, ``A novel bayesian similarity measure for
	recommender systems,'' in \emph{Proceedings of the 23rd International Joint
		Conference on Artificial Intelligence (IJCAI)}, 2013, pp. 2619--2625.
	
	\bibitem{majumder2019generating}
	\BIBentryALTinterwordspacing
	B.~P. Majumder, S.~Li, J.~Ni, and J.~McAuley, ``Generating personalized recipes
	from historical user preferences,'' in \emph{Proceedings of the 2019
		Conference on Empirical Methods in Natural Language Processing and the 9th
		International Joint Conference on Natural Language Processing
		(EMNLP-IJCNLP)}.\hskip 1em plus 0.5em minus 0.4em\relax Hong Kong, China:
	Association for Computational Linguistics, Nov. 2019, pp. 5976--5982.
	[Online]. Available: \url{https://aclanthology.org/D19-1613}
	\BIBentrySTDinterwordspacing
	
	\bibitem{Maxwell2015MovieLens}
	\BIBentryALTinterwordspacing
	F.~M. Harper and J.~A. Konstan, ``The movielens datasets: History and
	context,'' \emph{ACM Trans. Interact. Intell. Syst.}, vol.~5, no.~4, dec
	2015. [Online]. Available: \url{https://doi.org/10.1145/2827872}
	\BIBentrySTDinterwordspacing
	
	\bibitem{yahoo}
	\BIBentryALTinterwordspacing
	Webscope | yahoo labs | ratings and classification data. [Online]. Available:
	\url{https://webscope.sandbox.yahoo.com/catalog.php?datatype=r}
	\BIBentrySTDinterwordspacing
	
	\bibitem{portugal2017use}
	\BIBentryALTinterwordspacing
	I.~Portugal, P.~Alencar, and D.~Cowan, ``The use of machine learning algorithms
	in recommender systems: A systematic review,'' \emph{Expert Systems with
		Applications}, vol.~97, pp. 205--227, 2018. [Online]. Available:
	\url{http://www.sciencedirect.com/science/article/pii/S0957417417308333}
	\BIBentrySTDinterwordspacing
	
	\bibitem{Koren2008SVDPP}
	\BIBentryALTinterwordspacing
	Y.~Koren, ``Factorization meets the neighborhood: A multifaceted collaborative
	filtering model,'' in \emph{Proceedings of the 14th ACM SIGKDD International
		Conference on Knowledge Discovery and Data Mining}, ser. KDD '08.\hskip 1em
	plus 0.5em minus 0.4em\relax ACM, 2008, pp. 426--434. [Online]. Available:
	\url{http://doi.acm.org/10.1145/1401890.1401944}
	\BIBentrySTDinterwordspacing
	
	\bibitem{salakhutdinov2008PMF}
	\BIBentryALTinterwordspacing
	R.~Salakhutdinov and A.~Mnih, ``Probabilistic matrix factorization,'' in
	\emph{Proceedings of the 20th International Conference on Neural Information
		Processing Systems}, Curran Associates Inc.\hskip 1em plus 0.5em minus
	0.4em\relax Curran Associates, Inc., 2008, pp. 1257--1264. [Online].
	Available:
	\url{http://papers.nips.cc/paper/3208-probabilistic-matrix-factorization.pdf}
	\BIBentrySTDinterwordspacing
	
	\bibitem{koren2009BiasedMF}
	\BIBentryALTinterwordspacing
	Y.~Koren, R.~Bell, and C.~Volinsky, ``Matrix factorization techniques for
	recommender systems,'' \emph{Computer}, vol.~42, no.~8, pp. 30--37, 2009.
	[Online]. Available:
	\url{https://www.computer.org/csdl/magazine/co/2009/08/mco2009080030/13rRUxBa5fj}
	\BIBentrySTDinterwordspacing
	
	\bibitem{Hager2006Algorithm}
	\BIBentryALTinterwordspacing
	W.~W. Hager and H.~Zhang, ``Algorithm 851: Cg\_descent, a conjugate gradient
	method with guaranteed descent,'' \emph{ACM Transactions on Mathematical
		Software}, vol.~32, no.~1, p. 113–137, mar 2006. [Online]. Available:
	\url{https://doi.org/10.1145/1132973.1132979}
	\BIBentrySTDinterwordspacing
	
	\bibitem{Hager2005Anew}
	\BIBentryALTinterwordspacing
	------, ``A new conjugate gradient method with guaranteed descent and an
	efficient line search,'' \emph{SIAM Journal on Optimization}, vol.~16, no.~1,
	pp. 170--192, 2005. [Online]. Available:
	\url{https://doi.org/10.1137/030601880}
	\BIBentrySTDinterwordspacing
	
	\bibitem{guo2015librec}
	G.~Guo, J.~Zhang, Z.~Sun, and N.~Yorke-Smith, ``Librec: A java library for
	recommender systems.'' 2015.
	
\end{thebibliography}

\end{document}